\documentclass{iopart}
\usepackage{iopams}

\expandafter\let\csname equation*\endcsname\relax % these two lines
\expandafter\let\csname endequation*\endcsname\relax % allow loading mathtools
\usepackage{mathtools}

\usepackage[utf8]{inputenc}

\usepackage{tikz}
\usetikzlibrary{quantikz}

\usepackage{graphicx}
\usepackage{subfigure}
\graphicspath{ {figs/} } %% tidy subdir for figure files

\usepackage[numbers,sort&compress]{natbib}	
\setcitestyle{square,numbers}
\usepackage{xcolor}

\usepackage[breaklinks=true,%
            colorlinks=true,%
            linkcolor=blue,%
            urlcolor=blue,%
            citecolor=blue]{hyperref}
        
\usepackage{algpseudocode}
\usepackage{algorithm}

\renewcommand{\bra}[1]{\ensuremath{\left\langle#1\right|}}
\renewcommand{\ket}[1]{\ensuremath{\left|#1\right\rangle}}
\renewcommand{\braket}[2]{\ensuremath{\left\langle#1\middle\vert#2\right\rangle}}
\newcommand{\ketbra}[2]{\ensuremath{\left|#2\right\rangle\left\langle#1\right|}}

\newcommand{\pdiff}[2]{\ensuremath{\frac{\partial #1}{\partial #2}}}

\usepackage{bbold}

\begin{document}

\title{Improved maximum-likelihood quantum amplitude estimation}
\author{Adam Callison$^1$, Dan E. Browne$^1$}
\address{$^1$ Department of Physics and Astronomy, University College London, London, WC1E~6BT, UK}
\eads{\mailto{a.callison@ucl.ac.uk}}

\begin{abstract}
Quantum amplitude estimation is a key subroutine in a number of powerful quantum algorithms, including quantum-enhanced Monte Carlo simulation and quantum machine learning.
Maximum-likelihood quantum amplitude estimation (MLQAE) is one of a number of recent approaches that employ much simpler quantum circuits than the original algorithm based on quantum phase estimation. In this article, 
we deepen the analysis of MLQAE to put the algorithm in a more prescriptive form, including scenarios where quantum circuit depth is limited. In the process, we observe and explain particular ranges of `exceptional' values of the target amplitude for which the algorithm fails to achieve the desired precision.
We then propose and numerically validate a heuristic modification to the algorithm to overcome this problem, bringing the algorithm even closer to being useful as a practical subroutine on near- and mid-term quantum hardware.
\end{abstract}

\section{Introduction}

Quantum computing promises significant computational speedups compared to classical computing in a number of important tasks throughout science and industry \citep{montanaro2016quantum,cho2021quantum}, and has consequently been an active and important topic of research for several decades \citep{feynman1982simulating,benioff1980computer,deutsch1985quantum}.
Among the wide variety of quantum algorithms that have been developed exists a core set of general purpose subroutines with broad applicability, and a key example is the quantum amplitude estimation algorithm.
Quantum amplitude estimation can be used to estimate the probability that the output of another algorithm satisfies a particular target property, and can do this with a quadratic speedup in terms of the desired precision of the estimate, compared to classical sampling.
This can be employed as a subroutine in a number of important algorithms, including quantum-enhanced Monte Carlo simulation \citep{montanaro2015quantum,rebenstrost2018quantum,miyamoto2020reduction} and algorithms in quantum machine learning \citep{wiebe2015quantum,wiebe2016quantum,wiebe2016quantum2,li2019sublinear,kerenidis2019qmeans,miyahara2020quantum}.

The original quantum amplitude estimation due to Brassard et al. \citep{brassard2002quantum} essentially combines Grover's algorithm \citep{grover1996fast} and quantum phase estimation \citep{kitaev1995quantum}, and as a consequence involves quantum circuits with complicated controlled operations that may be challenging to implement in near- and mid-term quantum hardware.
More recently, a number of much simpler quantum amplitude estimation algorithms have been developed that employ much simpler circuits while achieving the same speedup \citep{aaronson2020quantum,venkateswaran2020quantum,nakaji2020faster,grinko2021iterative,rall2021faster,rall2022amplitude,manzano2022real,giurgicatiron2022low,fukuzawa2022modified}, including the maximum likelihood method (MLQAE) of Suzuki et al. \citep{suzuki2020amplitude} and its extensions \citep{brown2020quantum,tanaka2021amplitude,herbert2021noise,uno2021modified,giurgicatiron2022low,tanaka2022noisy,plekhanov2022variational}.

MLQAE employs a combination of Grover-like circuits and classical likelihood-maximisation-based post-processing in order to find a high-quality amplitude estimate, and as such can be viewed as a hybrid quantum-classical algorithm \citep{callison2022hybrid}.
Unlike the original quantum amplitude estimation algorithm and many of the other more recent proposals, the theoretical scaling analysis of maximum-likelihood quantum amplitude estimation relies on certain regularity and asymptotic assumptions related to the Bernstein-von Mises theorem \citep{vandervaart1998bernstein}, and as a consequence the quadratic speedup is not completely analytically rigorous.
Nevertheless, numerical experiments in prior work show that the algorithm appears to achieve the quadratic speedup in practice with low constant overhead using relatively simple quantum circuits.

In this work, we extend the analysis of MLQAE to give the algorithm a more prescriptive structure: given an achievable circuit depth (expressed abstractly as the achievable number of applications of a Grover-like operator), we show how to approximately target a particular desired precision with some target probability.
In the process of validating this extension through numerical simulations, we observe and explain in detail particular ranges of `exceptional' values of the target amplitude for which the algorithm fails to achieve the desired precision.
We then propose a heuristically-motivated modification to the algorithm to prevent this issue, and show through numerical simulation that our approach works.
Overall, through the insights in this work, we improve the understanding of MLQAE, and move it toward usefulness  as a practical algorithm in the near- and mid-term.

The article is structured as follows: in section \ref{sec:background}, we review the relevant background for quantum amplitude estimation, including the problem definition, a discussion of some of its applications, and a detailed description of maximum-likelihood quantum amplitude estimation and its information-theoretic analysis.
We begin our contribution in section \ref{sec:reqNshot} by extending the analysis and showing how to target a desired precision with a specified probability (up to certain regularity assumptions related to the Bernstein-von Mises theorem), including in depth-limited scenarios.
In doing so, we identify particular `exceptional' values of the target amplitude for which the algorithm fails to achieve the desired precision, and we explore and explain this phenomenon in detail in section \ref{sec:exceptionalvalues}.
In section \ref{sec:depthjitter}, we develop and numerically validate a heuristic method to mitigate the problematic exceptional values.
Finally, in section \ref{sec:summary}, we summarise the article and suggest some directions for future work.

\section{Background}
\label{sec:background}

\subsection{Problem definition}

The formal setting in which quantum amplitude estimation (QAE) is typically considered assumes that one has access to some unitary quantum algorithm $\hat{A}$ on $n$ qubits and an oracle $\hat{O}$ that partitions the computational basis $\{\ket{j}_n\}_{j=1}^{2^n}$ into two disjoint subsets, the `good' subset $G$ and the `bad' subset $B = \{\ket{j}_n\}_{j=1}^{2^n} - G$, via the action
\begin{align}
    \hat{O}\ket{j} = \left\{
    \begin{array}{lr}
        \ket{j}, & \mathrm{for\ }j \in G \\
        -\ket{j}, & \mathrm{for\ }j \in B.
    \end{array}\right.
\end{align}
The state prepared by applying the algorithm $\hat{A}$ to some reference state, which can be taken without loss of generality to be the all-zero state $\ket{0}^{\otimes n}$, can be written as
\begin{align}
    \ket{A}_n & \equiv \hat{A} \ket{0}^{\otimes n} \\
     & = \sqrt{a} \ket{A_G}_n + \sqrt{1-a} \ket{A_B}_n,
\end{align}
where the (normalised) states $\ket{A_G}_n$ and $\ket{A_B}_n$ have support only on the `good' and `bad' subsets respectively, with phases chosen such that $a$ is a positive real number.
With these definitions, the goal of QAE is to find an estimate $\tilde{a}$ of the probability $a$ (referred to in this setting as an amplitude) that is correct up to a particular precision $\epsilon$ with high probability.
The precision requirement is sometimes considered to be \textit{relative}
\begin{align}
    a(1 - \epsilon) \leq \tilde{a} \leq a(1 + \epsilon),
\end{align}
and is sometimes considered to be \textit{additive}
\begin{align}
    a - \epsilon \leq \tilde{a} \leq a + \epsilon.
\end{align}

\subsection{Conventional QAE}
\label{ssec:conventional_qae}

The first algorithm for quantum amplitude estimation is due to Brassard et al. \cite{brassard2002quantum}, and draws on ideas from Grover's algorithm \citep{grover1996fast} and the Quantum Phase Estimation (QPE) algorithm \citep{kitaev1995quantum}. 
The core of the algorithm is the Grover-like iteration operator 
\begin{align}
    \hat{Q} = \hat{A}\left(\mathbb{1}-2\ketbra{0}{0}_n\right)\hat{A}^\dagger \hat{O} \label{eq:groverQ},
\end{align}
which can be shown to have a pair of eigenvalues $\lambda_\pm = e^{\pm2i\theta_a}$, where $\theta_a$ is defined by
\begin{align}
    a \equiv \sin^2\theta_a \\
    0 \leq \theta_a \leq \frac{\pi}{2}.
\end{align}
The corresponding eigenvectors $\ket{\lambda_\pm}_n$ span the subspace $\mathrm{span}(\{\ket{A_G}_n, \ket{A_B}_n\})$ and have no support outside this subspace.

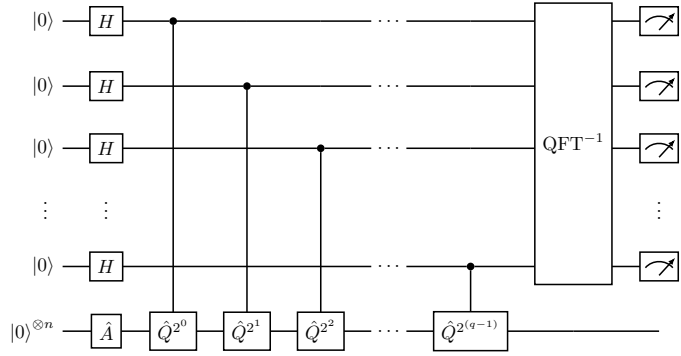
\begin{figure}
\begin{centering}
\resizebox{0.7\columnwidth}{!}{
\begin{quantikz}
\lstick{$\ket{0}$} & \gate{H} & \ctrl{5} & \qw & \qw & \ \ldots\ \qw & \qw & \gate[5, nwires=4]{\mathrm{QFT}^{-1}} & \meter{} \\
\lstick{$\ket{0}$} & \gate{H} & \qw & \ctrl{4} & \qw & \ \ldots\ \qw & \qw & & \meter{}\\
\lstick{$\ket{0}$} & \gate{H} & \qw & \qw & \ctrl{3} & \ \ldots\ \qw & \qw & & \meter{}\\
\lstick{\ \vdots\ } & \ \vdots\  & & & & & & & \ \vdots\ \\
\lstick{$\ket{0}$} & \gate{H} & \qw & \qw & \qw & \ \ldots\ \qw & \ctrl{1} & & \meter{}\\
\lstick{$\ket{0}^{\otimes n}$} & \gate{\hat{A}} & \gate{\hat{Q}^{2^0}} & \gate{\hat{Q}^{2^1}} & \gate{\hat{Q}^{2^2}} & \ \ldots\ \qw & \gate{\hat{Q}^{2^{(q-1)}}} & \qw & \qw \\
\end{quantikz}
}
\caption{\label{fig:brassardqae} 
Circuit for conventional quantum amplitude estimation
}
\end{centering}
\end{figure}

The original QAE algorithm uses the circuit shown in Fig. \ref{fig:brassardqae}; it requires not only the $n$-qubit (target) register needed for applying the algorithm $\hat{A}$, but an additional $q$-qubit \textit{control} register, used for controlling applications of the Grover iteration $\hat{Q}$.
The first application of the algorithm $\hat{A}$ on the target register prepares the state $\ket{A}_n$, which has no support outside $\mathrm{span}(\{\ket{A_G}_n, \ket{A_B}_n\})$, and thus no overlap with any eigenstates of the Grover iteration $\hat{Q}$ other than $\ket{\lambda_\pm}_n$.
The remainder of the circuit simply applies standard QPE to the Grover iteration $\hat{Q}$; that is, turning the results $m_0,m_1,\dots,m_q-1$ of the final layer of measurements into an integer $k=2^0m_0 + 2^1m_1+\dots,2^{q-1}m_{q-1}$ and converting it into an angle will, with high probability, result in either $\frac{\pi k}{2^q}=\theta_{\tilde{a}}$ or $\frac{\pi k}{2^q}=\pi - \theta_{\tilde{a}}$ where $|\theta_{\tilde{a}}- \theta_a|= O\left(\frac{1}{2^q}\right)$.
Thus, with high probability, the amplitude estimate
\begin{align}
    \tilde{a} &= \sin^2\left(\min\left[\left\{\frac{\pi k}{2^q}, \pi - \frac{\pi k}{2^q}\right\}\right]\right)
\end{align}
will be within an error $\epsilon \sim O\left(\frac{1}{2^q}\right)$ of the true amplitude $a$.

An application of the Grover iterator $\hat{Q}$ requires two calls to the algorithm $\hat{A}$ (more correctly, one call to $\hat{A}$ and one to its inverse $\hat{A}^\dagger$).
Thus, an application of a \textit{power} $l$ of $\hat{Q}$ ($\hat{Q}^l$) requires $2l$ applications of the algorithm $\hat{A}$.
Inspecting the target register in Fig. \ref{fig:brassardqae}, it can be seen that the total number of calls to $\hat{A}$ is
\begin{align}
    N_\mathrm{calls} &= 1 + 2\sum_{j=0}^{q-1} 2^j \\
    &= 2\cdot2^q - 1 \\
    &= O\left(\frac{1}{\epsilon}\right).
\end{align}
This constitutes a quadratic speedup over the case of simply sampling from the output of $\hat{A}$, for which the number of calls scales as $N_\mathrm{calls} = O\left(\frac{1}{\epsilon^2}\right)$.

While other forms of QPE exist that avoid the use of a quantum Fourier transform (QFT), they typically still require the unitary being studied (in this case the Grover iterator $\hat{Q}$) to be controlled \citep{kitaev1995quantum, higgins2009demonstrating, kimmel2015robust, wiebe2016efficient, knill2007optimal, dutkiewicz2021heisenberg}, or the ability to implement Haar-random unitaries \citep{zintchenko2016randomized}, both of which can be costly in terms of quantum resources \citep{suzuki2020amplitude}.

\subsection{Applications of quantum amplitude estimation}
In this subsection, we briefly review a number of applications of QAE as a subroutine in quantum algorithms.
\label{ssec:applications}
\subsubsection{Monte Carlo simulation}
\label{sssec:montecarlo}
In \citep{montanaro2015quantum}, Montanaro showed that QAE can be used as a subroutine in otherwise classical Monte Carlo algorithms in order to achieve a quadratic quantum speedup, compared to the best-known fully classical approaches.
As a simple example, consider the problem of estimating the mean $\mu$ of a randomized (classical) algorithm $A$ to within some additive error $\epsilon$ with high probability.
If an upper bound $\sigma^2$ is known on the variance of the random output from the algorithm $A$, then it is a well-known result from classical statistics that such an estimate $\tilde{\mu}$ for the mean $\mu$ requires a number of Monte Carlo samples that scales as
\begin{align}
    N_\mathrm{samples}^{(\mathrm{Monte\ Carlo})} &= O\left(\frac{\sigma^2}{\epsilon^2}\right).
\end{align}
Montanaro showed that, by essentially replacing the classical algorithm $A$ with a quantum algorithm $\hat{A}$ that outputs a quantum state which, if it were to be measured in the computational basis, would have the same probability distribution as the original algorithm $\hat{A}$, then QAE can be applied to produce an estimate $\tilde{\mu}$ of the mean $\mu$ more efficiently.
The number of calls to the quantum algorithm $\hat{A}$ in this case scales (up to polylogarithmic factors) as 
\begin{align}
    N_\mathrm{calls}^{(\mathrm{QAE\ Monte\ Carlo})} &= O\left(\frac{\sigma}{\epsilon}\right),
\end{align}
essentially a quadratic speedup over purely classical Monte Carlo.
As an important extension, he also pointed out that if the quantum algorithm $\hat{A}$ already has a quantum speedup compared to its classical counterpart $A$, that is if $\hat{A}$ produces `quantum samples' more efficiently than $A$ produces samples, then this speedup will \textit{concatenate} with the QAE Monte Carlo speedup.

Among a number of other more sophisticated examples, Montanaro showed how to use QAE to speed up Markov chain Monte Carlo algorithms for computing approximate partition functions; in this case, discrete-time quantum walks \citep{venegasandraca2012quantum} are used to gain a quadratic speedup in the mixing time of the Markov chain, which then concatenates with the QAE-related speedup, producing an overall speedup compared to the best known classical algorithms.
Other applications of quantum-enhanced Monte Carlo simulation are reported elsewhere \citep{rebenstrost2018quantum,miyamoto2020reduction}.

\subsubsection{Machine learning}
Within the flourishing field of quantum machine learning, two key research directions are that of quantum clustering algorithms and quantum classification algorithms \citep{wiebe2015quantum,kerenidis2019qmeans}.
In these algorithms, a core step is the estimation of the inner product 
 $\braket{\boldsymbol{x}}{\boldsymbol{y}}$  between two vectors  $\boldsymbol{x}$ and $\boldsymbol{y}$ for which unitary circuits $U_{\boldsymbol{x}}$ and $U_{\boldsymbol{y}}$ are known that can load them into quantum states via 
\begin{align}
    \ket{\boldsymbol{x}, \boldsymbol{y}} &= U_{\boldsymbol{x},\boldsymbol{y}}\ket{\boldsymbol{0}}.
\end{align}
By choosing the product of unitaries $U_{\boldsymbol{y}}^\dagger U_{\boldsymbol{x}}$ as the quantum algorithm $\hat{A}$ and the state $\ket{\boldsymbol{0}}$ as the (unique) `good' state (as well as the initial state), QAE can be performed to produce an estimate of the inner product
\begin{align}
    \braket{\boldsymbol{x}}{\boldsymbol{y}} &= \bra{\boldsymbol{0}}U_{\boldsymbol{y}}^\dagger U_{\boldsymbol{x}}\ket{\boldsymbol{0}},
\end{align}
as required.
Other applications of QAE within quantum machine learning are reported elsehwere \citep{wiebe2016quantum,wiebe2016quantum2,li2019sublinear,miyahara2020quantum}.

\subsubsection{Finance}
In the field of finance, an important task is to calculate a quantity known as the Economic Capital Requirement (ECR), which is essentially the amount of capital an investor must keep in order to protect against losses.
This in turn requires an accurate understanding of the risk associated with a particular portfolio, and an often used measure of this risk is the ``Value at Risk" (VaR) metric, defined for a loss distribution $\mathcal{L}$ as the smallest total loss with probability $P$ greater than or equal to the parameter $\alpha$; that is
 \begin{align}
     \mathrm{VaR}_\alpha\left[\mathcal{L}\right] &= \mathrm{inf}_{x\geq 0}\left[x|P(\mathcal{L}\leq x) \geq \alpha\right]
 \end{align}
where $\alpha$ is often chosen to be around 99.9\% \citep{egger2020quantum,egger2021credit}.
In \citep{egger2021credit}, Egger et al. showed that QAE can be used to gain a quantum speed-up for the calculation of VaR (and other related measures of risk).
To do this, Egger et al. propose a quantum algorithm $\hat{A}_\mathrm{VaR}$ that can be described at a high-level as being a combination of three parts; that is, $\hat{A}_\mathrm{VaR}=CSU$.
The first operator $U$ loads the correlated uncertainty for a portfolio of $k$ assets into one register of qubits.
The second operator $S$ then computes the total loss over the portfolio and stores the result, as a superposition, in another qubit register. 
The final operator $C$ applies a threshold to the total loss and flips a single target qubit if the total loss is less than or equal to a given $x$ value, and QAE is used to accurately determine the amplitude/probability of this target qubit having been flipped.
A classical outer loop is then used to perform a bisection search to find the smallest $x$ that is flipped with probability greater than or equal $\alpha$.

\subsection{QAE without QPE}

The conventional QAE of \cite{brassard2002quantum} (see subsection \ref{ssec:conventional_qae}) includes a number of complicated controlled operations, as well as a quantum Fourier transform as the final step (see Fig. \ref{fig:brassardqae}).
As a result, the entire procedure may be difficulty to implement, particularly on near- and mid-term quantum hardware.
The subsequent identification of a number of important applications for QAE (see subsection \ref{sssec:montecarlo}) led to a renewed interest in the algorithm, and a number of new approaches to the problem were developed that avoid the need for the quantum Fourier transform and the associated controlled-$\hat{Q}$ operations.

One of the first examples of QAE without QPE was proposed in \citep{aaronson2020quantum}.
In that work, Aaronson and Rall present a new, iterative algorithm for QAE that makes use of the Grover-like iteration operator $\hat{Q}$ but which does not use the quantum Fourier transform and does not require the $\hat{Q}$ operations to be controlled.
Furthermore, Aaronson and Rall gave a rigorous analysis, proving that the same quadratic speedup as in  conventional QAE is achieved, albeit with a large constant factor \citep{rall2022amplitude}.

Around the same time, an entirely different approach was proposed in \citep{suzuki2020amplitude} that, like the algorithm of \citep{aaronson2020quantum}, uses simple circuits consisting of various numbers of iterations of the Grover-like iteration operator $\hat{Q}$, but exploits the measurement outcomes via maximum-likelihood classical post-processing in order to make efficient use of the information gained in each measurement.
This maximum-likelihood QAE (MLQAE) also achieves the quadratic speedup of the conventional QAE in practice, and seemingly with a smaller constant factor than the algorithm of \citep{aaronson2020quantum}.
However, the analysis makes assumptions that are only approximately true, and thus the formal scaling results are themselves only approximate and without the rigor of previous QAE algorithms.
Many extensions to this algorithm have since been proposed, including a version intended for low-circuit-depth scenarios \citep{giurgicatiron2022low} (a different algorithm with more rigorous analysis, based on the Chinese Remainder Theorem, was also presented in this work) and versions which modify the algorithm and analysis to account for noise on the quantum device \citep{brown2020quantum,tanaka2021amplitude,herbert2021noise,uno2021modified,tanaka2022noisy,giurgicatiron2022low,giurgicatiron2022low2,plekhanov2022variational}.
Despite the formal drawbacks,  MLQAE  has the potential to be a promising algorithm in practice, and is the focus of the present work. We will detail the MLQAE algorithm in the next subsection.

There have also been other approaches to QAE that are iterative in style, similar to the algorithm of \citep{aaronson2020quantum}, which have reduced the constant factor on the scaling while retaining the rigorous guarantees \citep{wie2019simpler,venkateswaran2020quantum,nakaji2020faster,wang2021minimizing,grinko2021iterative,fukuzawa2022modified} (the algorithms of \citep{suzuki2020amplitude} and \citep{grinko2021iterative} are compared experimentally in \citep{rao2020quantum}), including one that is also sensitive to the sign of the amplitude \citep{manzano2022real}.
Other algorithms which have the added advantage of only requiring a single state preparation are presented in \citep{rall2021faster,rall2022amplitude}.

\subsection{Maximum-likelihood QAE}
\label{ssec:mlqae}
It has been shown that MLQAE \citep{suzuki2020amplitude} performs as well or better than many of the other QAE algorithms \citep{grinko2021iterative}. 
Due to this, MLQAE has the potential to be a promising QAE algorithm in practice, and so it is worthwhile to find ways to better understand and improve the MLQAE algorithm.

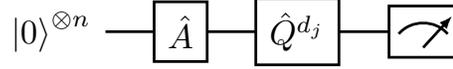
\begin{figure}
\begin{centering}
\resizebox{0.49\columnwidth}{!}{
\begin{quantikz}
\lstick{$\ket{0}^{\otimes n}$} & \gate{\hat{A}} & \gate{\hat{Q}^{d_j}} & \meter{} \\
\end{quantikz}
}
\caption{\label{fig:mlqae} 
Example circuit for maximum-likelihood quantum amplitude estimation.
}
\end{centering}
\end{figure}

MLQAE involves running a series of quantum circuits of the form shown in Fig. \ref{fig:mlqae}, much simpler than the circuit in Fig. \ref{fig:brassardqae}, and feeding the measurement outcomes into some classical post-processing that includes numerically maximising a likelihood function.
Like the conventional QAE algorithm of \cite{brassard2002quantum}, the MLQAE algorithm begins by applying the algorithm $\hat{A}$ to the $n$-qubit register (now the only the register) to prepare the state $\ket{A}$, and follows by applying some number of iterations of the Grover-like iterator $\hat{Q}$.
However, here, these applications of $\hat{Q}$ are not controlled on another register.

It can be shown that performing $d_j$ (where the index $j$ is introduced for later convenience) applications of the Grover iteration operator $\hat{Q}$ (which we will refer to as the `Grover-depth') to the state $\ket{A}=\sin\left(\theta_a\right)\ket{A_G} + \cos\left(\theta_a\right)\ket{A_G}$ produces the state
\begin{align}
    \ket{\psi_{d_j}} &\equiv \hat{Q}^{d_j}\ket{A} \\
    & = \sin\left[\left(2d_j+1\right)\theta_a\right]\ket{A_G} + \cos\left[\left(2d_j+1\right)\theta_a\right]\ket{A_B}.
    \label{eq:Qdj_action}
\end{align}
A computational-basis measurement then produces a `good' state from $G$ with probability
\begin{align}
    p_{d_j}\left(\theta_a\right) &= \sin^2\left[\left(2d_j+1\right)\theta_a\right].
    \label{eq:goodstateprob}
\end{align}
If the state $\ket{\psi_{d_j}}$ is prepared and measured some number $N_\mathrm{shot}$ times and the number of time a `good' state is found is denoted as $h_{d_j}$, then the likelihood $L_{d_j}(\theta_a=\theta;h_{d_j})$ of this value of $h_{d_j}$ as a function of the possible values $\theta$ for the angle $\theta_a$ is given by
\begin{align}
    L_{d_j}(\theta_a=\theta;h_{d_j}) &= \left[p_{d_j}\left(\theta\right)\right]^{h_{d_j}}\left[1-p_{d_j}\left(\theta\right)\right]^{N_\mathrm{shot}-h_{d_j}}.\label{eq:singledepth_lf}
\end{align}

In the MLQAE algorithm, a number of shots $N_\mathrm{shot}$ and a Grover-depth schedule $D=\left\{d_j\right\}_{j=0}^{q-1}$ are chosen (where the number of depths $q$ can be considered arbitrary at this point), and for each depth $d_j$ a circuit like the one in Fig. \ref{fig:mlqae} is used to prepare and measure the $\ket{\psi_{d_j}}$ state $N_\mathrm{shot}$ times, and the measurement record $\boldsymbol{h}=\left(h_{d_0}, h_{d_1}, \dots, h_{q-1}\right)$ is obtained.
The individual likelihood functions [Eq. \eqref{eq:singledepth_lf}] can then be combined to produce an overall likelihood function for the entire measurement record $\boldsymbol{h}$
\begin{align}
    L(\theta_a=\theta;\boldsymbol{h}) &= \prod_{j=0}^{q-1}L_{d_j}(\theta_a=\theta;h_{d_j})
    \label{eq:lf}
\end{align}

\begin{figure}
\begin{centering}
\subfigure[]{\includegraphics[width=0.49\columnwidth]{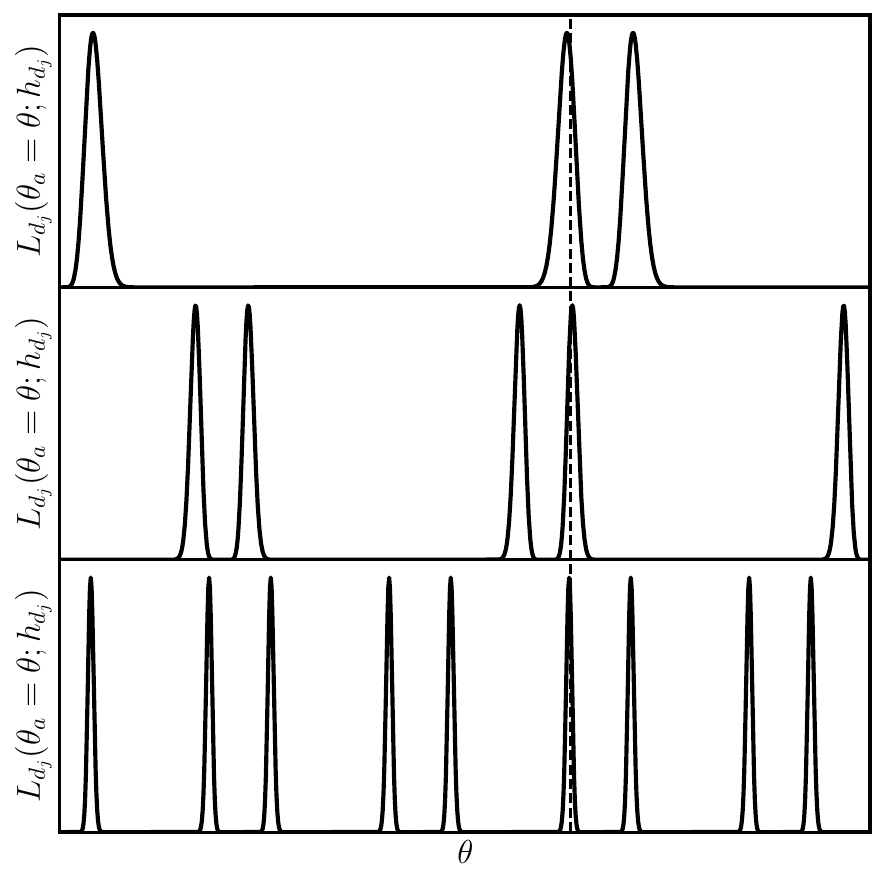}\label{fig:mlqae_likelihooda}}
\subfigure[]{\raisebox{0.88\height}{\includegraphics[width=0.49\columnwidth]{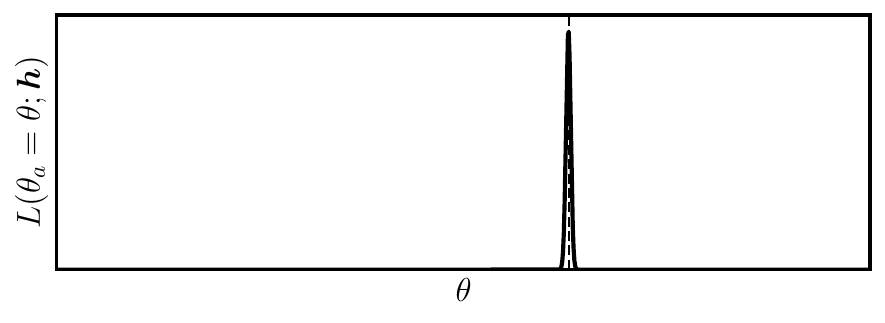}}\label{fig:mlqae_likelihoodb}}
\caption{\label{fig:mlqae_likelihood} 
(a) Illustrations of the likelihood $L_{d_j}(\theta_a=\theta;h_{d_j})$ that the angle $\theta_a$ is equal to a value $\theta$, given that a Grover-like QAE circuit with $d_j$ Grover iterations ($d_j$ larger for lower plots) produced $h_{d_j}$ `good' states from $N_\mathrm{shot}$ shots.
The actual angle $\theta_a$ that generated the outcomes is indicated by the dashed line.
(b) The overall likelihood $L(\theta_a=\theta;\boldsymbol{h})$ produced by multiplying together the likelihoods for each $d_j$.
}
\end{centering}
\end{figure}

Typical forms for the individual likelihood functions $L_{d_j}(\theta_a=\theta;h_{d_j})$ are illustrated in Fig. \ref{fig:mlqae_likelihooda} ($d_j$ larger for lower plots).
It can be seen that smaller Grover-depths $d_j$ produce a likelihood function with few broad peaks, while larger Grover-depths produce a likelihood function with many sharp peaks.
The typical form of the overall likelihood function, produced by multiplying together the likelihoods for each $d_j$, is illustrated in Fig. \ref{fig:mlqae_likelihoodb}; the combined effect is a single, sharp peak which, if located by a numerical maximisation, will produce an accurate estimate of $\theta_{\tilde{a}}$ of the angle $\theta_a$ (which can be turned into an accurate estimate $\tilde{a}$ for the amplitude $a$).

\subsubsection{Information theory analysis of MLQAE}
\label{sssec:infotheory_mlqae}
It can be shown (see appendix \ref{secapp:fisherinfoproof}) that the Fisher information carried by the measurement record $\boldsymbol{h}$ regarding the amplitude $a$ is
\begin{align}
     I(a) &= \frac{N_\mathrm{shot}}{a(1-a)}\sum_{j=0}^{q-1}(2d_j + 1)^2 \label{eq:fisher_info_predef}\\
     &= \frac{N_\mathrm{shot}}{a(1-a)}(S_D^{(2)})^2, \label{eq:fisher_info}
\end{align}
where in the second line the definition 
\begin{align}
     S_D^{(2)} &\equiv \sqrt{\sum_{j=0}^{q-1}(2d_j + 1)^2}
     \label{eq:SD2}
\end{align}
has been made.

Assuming the likelihood function satisfies certain regularity conditions detailed in the Bernstein–von Mises theorem \citep{vandervaart1998bernstein}, which will be explained in more detail in section \ref{sec:exceptionalvalues}, then as $N_\mathrm{shot}$ increases the estimate $\tilde{a}$ becomes approximately distributed according to a normal distribution $\mathcal{N}\left[a, I^{-1}(a)\right]$ with mean equal to the true amplitude $a$ and standard deviation related to the Fisher information as $I^{-\frac{1}{2}}(a)$.
That is, the average additive error in the estimate $\tilde{a}$ will be approximately
\begin{align}
    \epsilon_\mathrm{avg} &\equiv \sqrt{\mathbb{E}\left[(\tilde{a}-a)^2\right]} \nonumber \\
     &\approx \sqrt{\frac{a(1-a)}{N_\mathrm{shot}}}\cdot\frac{1}{S_D^{(2)}}.  \label{eq:mlqae_avgerror_from_Nshot}
\end{align}
The number of calls to the algorithm $\hat{A}$, including the call at the beginning of each circuit and the two calls required for each application of the Grover iteration operator $\hat{Q}$, is
\begin{align}
    N_\mathrm{calls} &= N_\mathrm{shot}\sum_{j=0}^{q-1} \left(2d_j + 1\right) \\
    &= N_\mathrm{shot}S_D^{(1)}
    \label{eq:Ncalls_from_Nshot}
\end{align}
where in the second line the definition 
\begin{align}
     S_D^{(1)} &\equiv \sum_{j=0}^{q-1}(2d_j + 1)
     \label{eq:SD1}
\end{align}
has been made.
Combining Eq. \eqref{eq:mlqae_avgerror_from_Nshot} and Eq. \eqref{eq:Ncalls_from_Nshot}
gives
\begin{align}
     \epsilon_\mathrm{avg} &\approx \sqrt{\frac{a(1-a)}{N_\mathrm{calls}}}\cdot\frac{\sqrt{S_D^{(1)}}}{S_D^{(2)}}.\label{eq:mlqae_avgerror_from_Ncalls}
\end{align}

The primary Grover-depth schedule considered by Suzuki et al. in \cite{suzuki2020amplitude} is the \textit{exponential} schedule
\begin{align}
    D_\mathrm{EXP} &= \left\{d_0 = 0\right\}\cup \left\{d_j = 2^{j-1}\right\}_{j=1}^{q-1}.
    \label{eq:expsched}
\end{align}
With this choice, and writing the maximum Grover-depth as $d\equiv d_{q-2}$, the quantities $S_D^{(1,2)}$ can be evaluated (except for the trivial case of $d=0$) as
\begin{align}
    S_{D_\mathrm{EXP}}^{(1)} &= 4d + \log_2d \label{eq:SD1exp} \\
    &= O(d) \label{eq:SD1exp_scaling} \\
    S_{D_\mathrm{EXP}}^{(2)} &= \sqrt{\frac{16d^2}{3} + 8d + \log_2 d - \frac{10}{3}} \label{eq:SD2exp}\\
    &= O(d). \label{eq:SD2exp_scaling}
\end{align}
Combining the scalings in Eq. \eqref{eq:SD1exp_scaling} and Eq. \eqref{eq:SD2exp_scaling} with Eq. \eqref{eq:mlqae_avgerror_from_Ncalls} gives
\begin{align}
    \epsilon_\mathrm{avg}\sqrt{N_\mathrm{calls}} & \approx O\left(\frac{1}{\sqrt{d}}\right). \label{eq:epsilonscaling}
\end{align}
Thus, the quadratic speedup of $\epsilon_\mathrm{avg} \approx O\left(N_\mathrm{calls}^{-1}\right)$ over simply sampling from the output of the algorithm $\hat{A}$ can be achieved by allowing (via the choice of the number $q$ of depths) the maximum Grover-depth $d$ to scale linearly with the desired average accuracy as $d = O\left(\epsilon_\mathrm{avg}^{-1}\right)$.
The numerical maximisation of the likelihood function can also be performed in time $O\left(\epsilon_\mathrm{avg}^{-1}\right)$ by simply calculating the likelihood function (or, in practice, its logarithm) at $O\left(\epsilon_\mathrm{avg}^{-1}\right)$ regularly-spaced values of $\theta$ and choosing the maximum, though this may not be the fastest method in practice.

It is worth noting that in the case where the maximum Grover-depth $d$ scales with the desired average accuracy as $d = O\left(\epsilon_\mathrm{avg}^{-1}\right)$, the average error $\epsilon_\mathrm{avg}$ and the target amplitude $a$ would appear in Eq. \eqref{eq:mlqae_avgerror_from_Ncalls} at the same order, suggesting that a version of this algorithm exists that achieves \textit{relative} error. 
More explicitly, we find from Eq. \eqref{eq:mlqae_avgerror_from_Ncalls} that 
\begin{align}
     \frac{\epsilon_\mathrm{avg}^2}{a(1-a)}N_\mathrm{calls} &\approx \frac{S_D^{(1)}}{\left(S_D^{(2)}\right)^2} \nonumber \\
     &= O\left(\frac{1}{d}\right) \nonumber \\
     &= O\left(\epsilon_\mathrm{avg}\right) \nonumber \\
     \frac{\epsilon_\mathrm{avg}}{a(1-a)}N_\mathrm{calls} &\approx O(1). \nonumber
\end{align}
Making use of this scaling in practice will require a more sophisticated approach to the numerical likelihood maximisation, such as a grid-search that iteratively focuses more finely on smaller regions of the $\theta$ domain, to ensure that it can be performed in time $O\left(\frac{a(1-a)}{\epsilon_\mathrm{avg}}\right)$ rather than in time  $O\left(\epsilon_\mathrm{avg}^{-1}\right)$.
While this is an interesting observation, we will not consider relative error further in this work.

\section{Targeting a desired precision}
\label{sec:reqNshot}
Suzuki et al.'s analysis \citep{suzuki2020amplitude} shows how the maximum Grover-depth $d$ can be scaled to achieve a quadratic speedup compared to classical sampling, but leaves the number of shots $N_\mathrm{shot}$ performed at each Grover-depth as a free parameter.
In this section, we extend the analysis to show how, for a given maximum Grover-depth $d$, the number of shots $N_\mathrm{shot}$ can be chosen to approximately target a particular precision $\epsilon$ with probability at least $1-\delta$.

As mentioned in subsection \ref{ssec:mlqae}, the Bernstein-von Mises theorem implies, assuming the likelihood function satisfies its regularity conditions, that for sufficiently large number of shots $N_\mathrm{shot}$, the estimate $\tilde{a}$ becomes approximately distributed according to a normal distribution $\mathcal{N}\left[a, I^{-1}(a)\right]$ with mean equal to the true amplitude $a$ and standard deviation related to the Fisher information as $I^{-\frac{1}{2}}(a)$.
The probability that the estimate $\tilde{a}$ is within a precision $\epsilon$ of the true amplitude $a$ is thus approximately
\begin{align}
    P(|\tilde{a}-a| \leq \epsilon) & \approx \frac{1}{I^{-\frac{1}{2}}(a)\sqrt{2\pi}}\intop_{a-\epsilon}^{a+\epsilon}\mathrm{d}x \exp\left[-\frac{(x-a)^2}{2 I^{-1}(a)}\right].
\end{align}
Writing this probability as $1-\delta \equiv P(|\tilde{a}-a| \leq \epsilon)$, it can be shown that
\begin{align}
    I(a) \approx \frac{2\mathrm{erfinv}^2\left(1-\delta \right)}{\epsilon^2},
    \label{eq:fisher_info_from_eps_delta}
\end{align}
where $\mathrm{erfinv}$ is the inverse error function.
Combining Eq. \ref{eq:fisher_info_from_eps_delta} and Eq. \ref{eq:fisher_info} then gives an expression for the required number of shots
\begin{align}
    N_\mathrm{shot} \approx a(1-a)\frac{2\mathrm{erfinv}^2\left(1-\delta \right)}{\left(S_D^{(2)}\right)^2\epsilon^2}.
    \label{eq:req_Nshot_inc_a}
\end{align}
Given that the goal is to find an estimate for the amplitude $a$,  which is not known \textit{a priori}, we must also maximise $N_\mathrm{shot}$ over all values of the amplitude $a$.
This maximum occurs at $a=0.5$ and gives
\begin{align}
    N_\mathrm{shot} \approx \frac{\mathrm{erfinv}^2\left(1-\delta \right)}{2\left(S_D^{(2)}\right)^2\epsilon^2}.
    \label{eq:req_Nshot}
\end{align}
Hereafter, we will focus on the regime in which the targeted precision $\epsilon$ and acceptable failure probability $\delta$ are small compared to $1/\left(S_D^{(2)}\right)^2=O(1/d^2)$ to avoid contradicting the the assumption that the number of shots is sufficiently large for the Bernstein-von Mises theorem to apply.
For small $\delta$, the $\mathrm{erfinv}^2\left(1-\delta \right)$ factor can be approximated as $\ln\left(\frac{\sqrt{\pi}}{\delta}\right)$ \citep{blair1976rational}.
Returning to the exponential schedule $D_\mathrm{EXP}$ from Eq. \eqref{eq:expsched} and relating $N_\mathrm{shot}$ to $N_\mathrm{calls}$ by Eq. \eqref{eq:Ncalls_from_Nshot}, this gives a scaling of
\begin{align}
    \epsilon\sqrt{N_\mathrm{calls}} & \approx O\left(\sqrt{\frac{\ln\left(\frac{1}{\delta}\right)}{d}}\right),
    \label{eq:epsilondeltascaling}
\end{align}
in agreement with Eq. \eqref{eq:epsilonscaling}.

\subsection{Depth-limited scenarios}
\label{ssec:depthlimited}

Relevant to implementations of MLQAE on near- and mid-term hardware, it is important to consider scenarios in which the maximum Grover-depth $d$ cannot be made to scale linearly with the desired precision $\epsilon$.
In prior work  \citep{giurgicatiron2022low}, Giurgica-Tiron et al. introduce a parameter $0\leq \beta \leq 1$, and allow the maximum Grover-depth $d$ to scale only as $O\left(\epsilon^{-(1-\beta)}\right)$.
Then, a polynomial Grover-depth schedule is chosen as
\begin{align}
    D_{\mathrm{POLY},\beta} &= \left\{d_j = \mathrm{Round}\left(j^{\frac{1-\beta}{2\beta}}\right)\right\}_{j=1}^{q},
\end{align}
where $q=\max\left(\frac{1}{\epsilon^{2\beta}}, \ln\left(\frac{1}{\epsilon}\right)\right)$ so that the schedule approaches an exponential schedule for small $\beta$ and the maximum Grover-depth $d \equiv d_q$ becomes approximately $\frac{1}{\epsilon^{1-\beta}}$ otherwise, as required.
Through analysis of the Fisher information content of measurements, Giurgica-Tiron et al. show that that the total number of calls $N_\mathrm{calls}$ to the algorithm $\hat{A}$ scales with the precision $\epsilon$ as 
\begin{align}
    N_\mathrm{calls} & \approx O\left(\frac{1}{\epsilon^{1+\beta}}\right).
    \label{eq:kpscaling}
\end{align}
From this result, it can be seen that this scheme allows for trade-off between circuit depth and quantum speedup compared to classically sampling from simple measurements of the algorithm $\hat{A}$.

However, we observe here that the same trade-off can be achieved using the standard exponential schedule $D_\mathrm{EXP}$ (or a similar one, as will be explained below), by simply choosing the maximum Grover-depth $d$ to scale appropriately as $O\left(\epsilon^{-(1-\beta)}\right)$.
In this case, it can be seen from Eq. \eqref{eq:epsilondeltascaling} (or even Eq. \eqref{eq:epsilonscaling}) that the number of calls $N_\mathrm{calls}$ scales in the same way as in Eq. \eqref{eq:kpscaling}.
If the Grover-depth scaling of $d=O\left(\epsilon^{-(1-\beta)}\right)$ is not compatible with $D_\mathrm{EXP}$ as defined in Eq. \eqref{eq:expsched}, then once a maximum Grover-depth $d$ is identified for a particular run of MLQAE, a similar schedule can be chosen as
\begin{align}
    D_{\mathrm{EXP},\nu} &= \left\{d_0 = 0\right\}\cup \left\{d_j = \mathrm{Round}\left(\nu^{j-1}\right)\right\}_{j=1}^{q-1},
    \label{eq:expschednu}
\end{align}
where the base $\nu$ and number of depths $q$ are chosen together such that $\nu$ is the closest value to $2$ that supports $\nu^{q-2}=d$.
Explicit pseudo-code for generating the schedule is given in Algorithm \ref{alg:limiteddepth}.
\begin{figure}[htb]
    \centering
    \begin{minipage}{0.89\textwidth}
        \begin{algorithm}[H]
                \caption{Generate depth-limited exponential schedule\label{alg:limiteddepth}}
                \textbf{Input}: $d$: Maximum Grover-depth. \newline 
                \textbf{Output}: $D_{\mathrm{EXP},\nu}$: Compatible exponential Grover-depth schedule, in ascending order.
                \begin{algorithmic}[1]
                \State \textbf{set} $p_{\max,\mathrm{ceil}} \coloneqq \mathrm{Ceil}[\log_2 d]$, $p_{\max,\mathrm{floor}} \coloneqq \mathrm{Floor}[\log_2 d]$ \textcolor{gray}{\texttt{\#two options for the maximum power}}
                \State \textbf{if} $p_{\max,\mathrm{floor}} = 0$: \textcolor{gray}{\texttt{\#discard a maximum power of 0}}
                \State \hspace{20pt} \textbf{set} $p_{\max} \coloneqq p_{\max,\mathrm{ceil}}$
                \State \hspace{20pt} \textbf{set} $\nu \coloneqq d^{1/p_{\max}}$
                \State \textbf{else}:
                \State \hspace{20pt} \textbf{set} $\mathtt{base\_upper} \coloneqq  d^{1/p_{\max,\mathrm{ceil}}}$, $\mathtt{base\_lower} \coloneqq d^{1/p_{\max,\mathrm{floor}}}$
                \State \hspace{20pt} \textbf{if} $|\mathtt{base\_upper}-2| < |\mathtt{base\_lower}-2|$:\textcolor{gray}{\texttt{\#choose the base closest to 2}}
                \State \hspace{40pt} \textbf{set} $p_{\max} \coloneqq p_{\max,\mathrm{ceil}}$, $\nu \coloneqq \mathtt{base\_upper}$
                \State \hspace{20pt} \textbf{else}:
                \State \hspace{40pt} \textbf{set} $p_{\max} \coloneqq p_{\max,\mathrm{floor}}$, $\nu \coloneqq \mathtt{base\_lower}$
                \State \textbf{set} $D_{\mathrm{EXP},\nu} = \left\{d_0 = 0\right\}\cup \left\{d_j = \mathrm{Round}\left(\nu^{j}\right)\right\}_{j=0}^{p_{\max}}$
                \State \textbf{return} $D_{\mathrm{EXP},\nu}$
                \end{algorithmic}
        \end{algorithm}
    \end{minipage}
\end{figure}
The difference $|\nu-2|$ becomes smaller over larger scales of maximum Grover-depth $d$ (see appendix \ref{secapp:nubase}), meaning that the approximate scaling results for the standard exponential schedule $D_\mathrm{EXP}$ shown in subsubsection \ref{sssec:infotheory_mlqae} and in this section still hold approximately.

An important case to consider is that of $\beta=1$; that is, when the maximum Grover-depth $d$ is fixed and not allowed to scale with the desired precision $\epsilon$.
This case is relevant to near- and mid-term hardware for which the useful circuit depth, and thus the achievable Grover-depth $d$ for MLQAE on a particular algorithm $\hat{A}$, will have some upper-limit on any particular device due to interactions with the environment.
Indeed, even specific devices employing error-correction will have some limit on the achievable Grover-depth $d$, as the available number of qubits will limit the size of error-correcting codes \citep{babbush2021focus}.
In this case, as the maximum Grover-depth $d$ is constant, it is clear from the approximate scaling in Eq. \eqref{eq:epsilondeltascaling} that the required number of calls $N_\mathrm{calls}$ will scale inversely-proportionally to the \textit{square} $\epsilon^2$ of the desired precision, as in the case of classical sampling; that is, no scaling advantage is achieved.
However, it is also the case in this case that the achievable Grover-depth $d$ determines the constant factor on this scaling.
Combining Eq. \eqref{eq:Ncalls_from_Nshot} and Eq. \eqref{eq:req_Nshot} gives the required number of calls $N_\mathrm{calls}$ to be
\begin{align}
    N_\mathrm{calls} \approx \frac{S_{D_{\mathrm{EXP},\nu}}^{(1)}}{\left(S_{D_{\mathrm{EXP},\nu}}^{(2)}\right)^2}\cdot\frac{\mathrm{erfinv}^2\left(1-\delta \right)}{2\epsilon^2}.
\end{align}
The speed-up factor $R_d\equiv N_\mathrm{calls}^{(\mathrm{classical})}/N_\mathrm{calls}$ achieved by using a maximum Grover-depth $d$ compared to classical sampling case (for which $d=0$ and $S_D^{(1)}=S_D^{(2)}=1$), can be quantified approximately as
\begin{align}
    R_d & \approx \frac{\left(S_{D_{\mathrm{EXP},\nu}}^{(2)}\right)^2}{S_{D_{\mathrm{EXP},\nu}}^{(1)}}.
\end{align}
Taking for simplicity the case where the maximum Grover-depth $d$ is a power of $2$, this can be rewritten as 
\begin{align}
    R_d & \approx \frac{\frac{16d^2}{3} + 8d + \log_2 d - \frac{10}{3}}{4d + \log_2d} \\
    &= O(d). \label{eq:speedupfactorscaling}
\end{align}

The meaning of Eq. \eqref{eq:speedupfactorscaling} is that, assuming the Bernstein-von Mises theorem holds, then using a quantum processor that supports a maximum Grover-depth $d$ (for the particular algorithm $\hat{A}$) to run MLQAE should produce an estimate $\tilde{a}$ for the amplitude $a$ within precision $\epsilon$ with probability approximately greater than $1-\delta$ with a $O(d)$ factor speed-up compared to simply classically sampling from the algorithm $\hat{A}$.
If we return to the case where the maximum Grover-depth $d$ can scale with the targeted precision $\epsilon$ as $d=O\left(\epsilon^{-1}\right)$, then a quadratic speedup over the classical scaling of $N_\mathrm{calls}^{(\mathrm{classical})}=O(\epsilon^{-2})$ is achieved, as expected.

In practice, even when limiting the circuit-depth to keep the noise under control, low levels of noise may still affect the results and place a limit on the achievable precision $\epsilon$.
Methods to mitigate this issue by modifying the the likelihood functions $L_{d_j}(\theta=\theta_a;h_{d_j})$ to include the effects of noise are discussed in \citep{suzuki2020amplitude,brown2020quantum,tanaka2021amplitude,giurgicatiron2022low,tanaka2022noisy}, while other QAE-specific noise-mitigation methods are discussed in \citep{herbert2021noise, uno2021modified,giurgicatiron2022low2,plekhanov2022variational}.
In this work, we do not consider the effects of noise beyond the notion of a depth limitation. 

\subsection{Numerical validation}
\label{ssec:numval}

To test the analysis presented in this section, we have performed numerical simulations of the MLQAE algorithm.
Given the simple description of the action of the operators $\hat{Q}^{d_j}$ in terms of the angle $\theta_a$ given in Eq. \ref{eq:Qdj_action}, it is not necessary to perform full quantum-circuit simulations in order to simulate MLQAE  for a particular pre-chosen value of the amplitude $a$.
Instead, for each Grover-depth $d_j$ in the chosen schedule $D$, we calculate the `good' state probability $p_{dj}\left(\theta_a\right)$, according to Eq. \eqref{eq:goodstateprob}, draw a number of `good' states $h_{d_j}$ from the binomial distribution $\mathcal{B}\left[N_\mathrm{shot}, p_{d_j}\left(\theta_a\right)\right]$, and collect these together into the measurement record $\boldsymbol{h}$.
From this measurement record, we construct the likelihood function $L(\theta_a=\theta;\boldsymbol{h})$ according to Eq. \eqref{eq:lf} and numerically maximise its logarithm by simply calculating it at $~3/\epsilon$ regularly-spaced values of $\theta$ (where the 3 has been chosen arbitrarily) and finding the maximum.

\begin{figure}
\begin{centering}
\subfigure[]{\includegraphics[width=0.49\columnwidth]{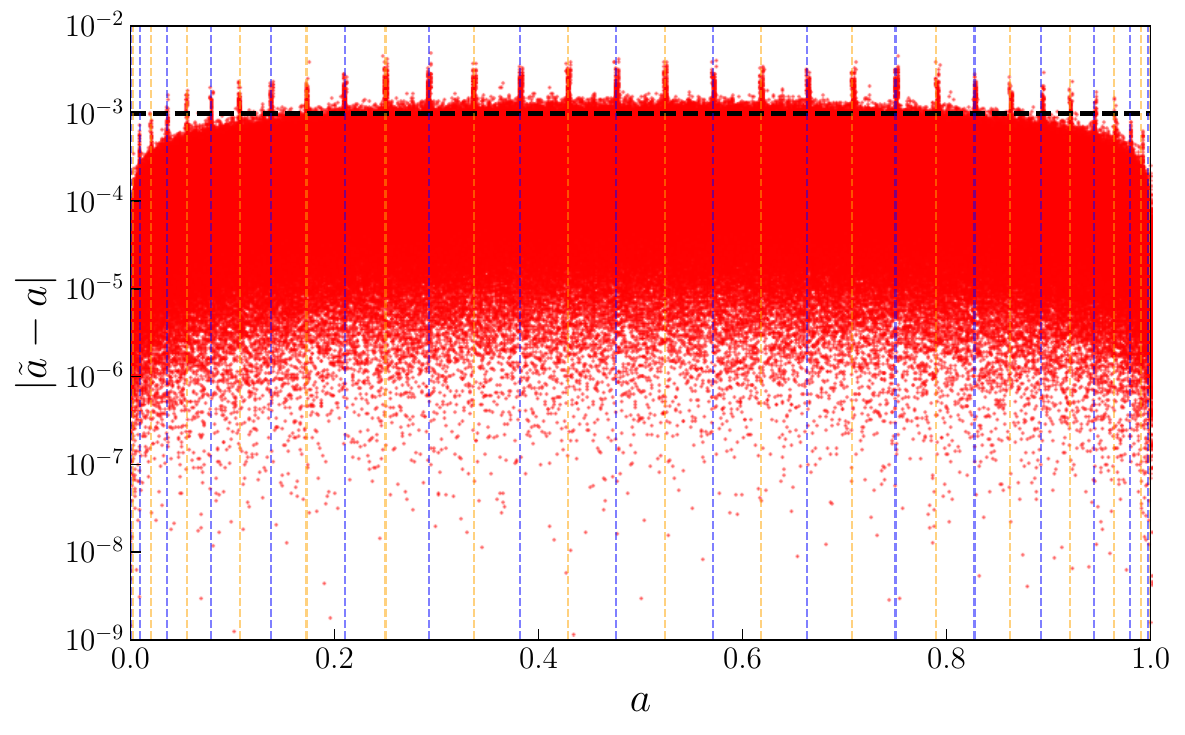}\label{fig:scatter16}}
\subfigure[]{\includegraphics[width=0.49\columnwidth]{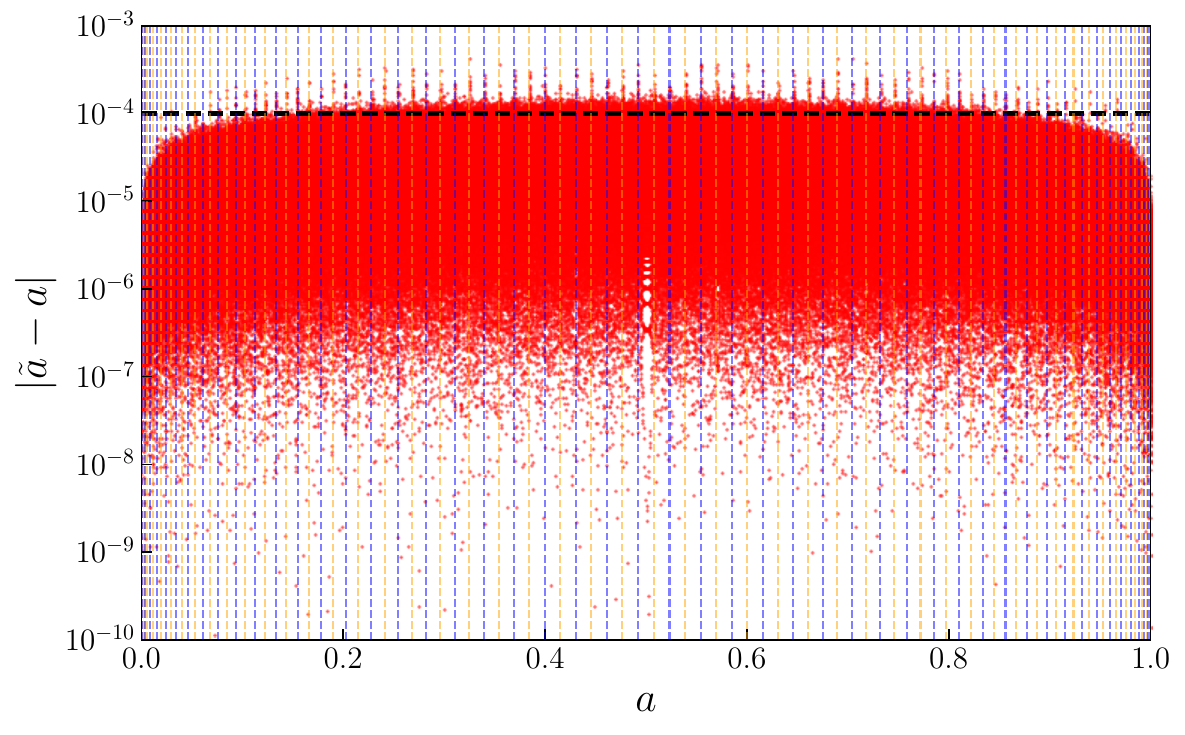}\label{fig:scatter50}}
\caption{\label{fig:scatter} 
The error in an estimate $\tilde{a}$, produced from a single run of the MLQAE algorithm, of the amplitude $a$ for $1\,000\,000$ equally-spaced values of the amplitude $a$ covering the full range from $a=0$ to $a=1$, using maximum Grover-depth $d$, targeting a precision $\epsilon$ with failure probability at most $\delta$ and a number of shots $N_\mathrm{shot}$ calculated according to Eq. \eqref{eq:req_Nshot}.
The black, horizontal dashed line is the target precision $\epsilon$, while the vertical dashed lines indicate the exceptional values according to Eq. \eqref{eq:exceptional_values}, where the blue (orange) lines are fore even (odd) $j$.
The parameters for the two plots are (a) $d=16$, $\epsilon=10^{-3}$, $\delta=0.01$ (resulting in $N_\mathrm{shot}=1111$) and (b) $d=50$, $\epsilon=10^{-4}$, $\delta=0.01$ (resulting in $N_\mathrm{shot}=11\,688$).
}
\end{centering}
\end{figure}

Fig. \ref{fig:scatter} shows, for two sets of parameters, the error in the estimate $\tilde{a}$, produced from a single run of the MLQAE algorithm, of the amplitude $a$ for $1\,000\,000$ equally-spaced values of the amplitude $a$ covering the full range from $a=0$ to $a=1$, targeting a precision $\epsilon$ with failure probability at most $\delta$, and using using maximum Grover-depth $d$ and a number of shots $N_\mathrm{shot}$ calculated according to Eq. \eqref{eq:req_Nshot}.
The parameters for the two plots are (a) $d=16$, $\epsilon=10^{-3}$, $\delta=0.01$ (resulting in $N_\mathrm{shot}=1111$) and (b) $d=50$, $\epsilon=10^{-4}$, $\delta=0.01$ (resulting in $N_\mathrm{shot}=11\,688$).

It can be seen in both plots of Fig. \ref{fig:scatter} that, for most typical values of the amplitude $a$, MLQAE produces an estimate with an error that is either close to or much smaller than the target precision $\epsilon$.
However, it is also clear that some small, regularly-spaced regions of the amplitude $a$ exist in which the estimates $\tilde{a}$ produced are significantly worse.
These exceptional values occur in the vicinity of $a$-values for which the circuit with the maximum Grover-depth $d$ produces `good' states with probability $1$ or $0$; that is, the exceptional values occur in the vicinity of 
\begin{align}
    a^{(d\rightarrow p_d(\theta_a)=0,1)}_k & = \sin\left[\frac{k\pi}{2\left(2d+1\right)}\right]^2\mathrm{\ for\ } k=0,1,\dots,2d,2d+1.
    \label{eq:exceptional_values}
\end{align}
The reason for this, as well as precisely what is meant by ``in the vicinity of", will be explained in detail in the next section, but for now, this observation allows us to examine the exceptional values more closely.

\begin{figure}
\begin{centering}
\subfigure[]{\includegraphics[width=0.49\columnwidth]{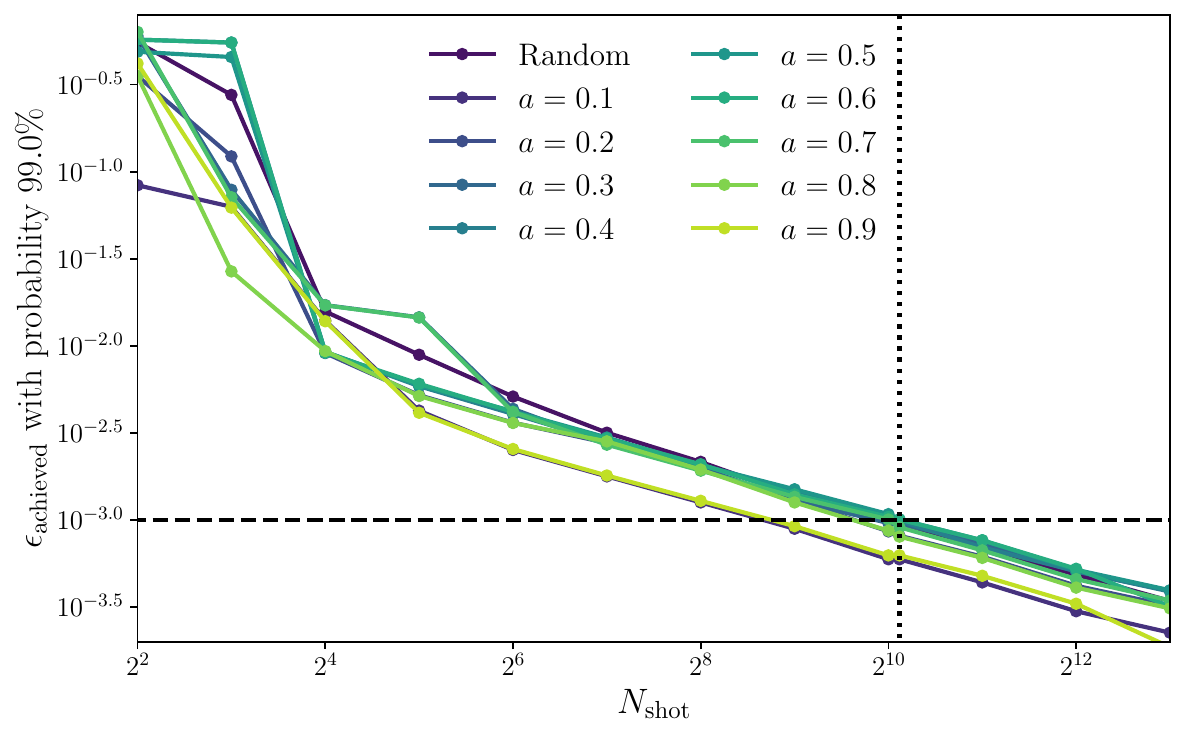}\label{fig:normalvalues16}}
\subfigure[]{\includegraphics[width=0.49\columnwidth]{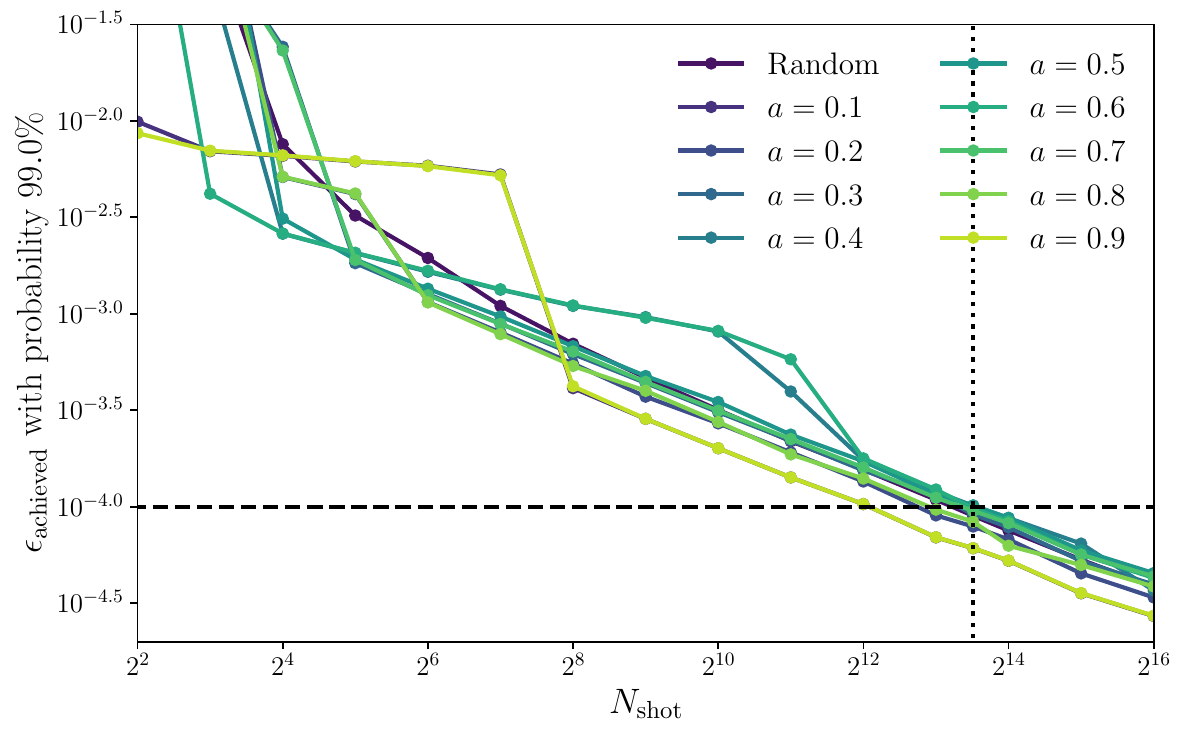}\label{fig:normalvalues50}}
\subfigure[]{\includegraphics[width=0.49\columnwidth]{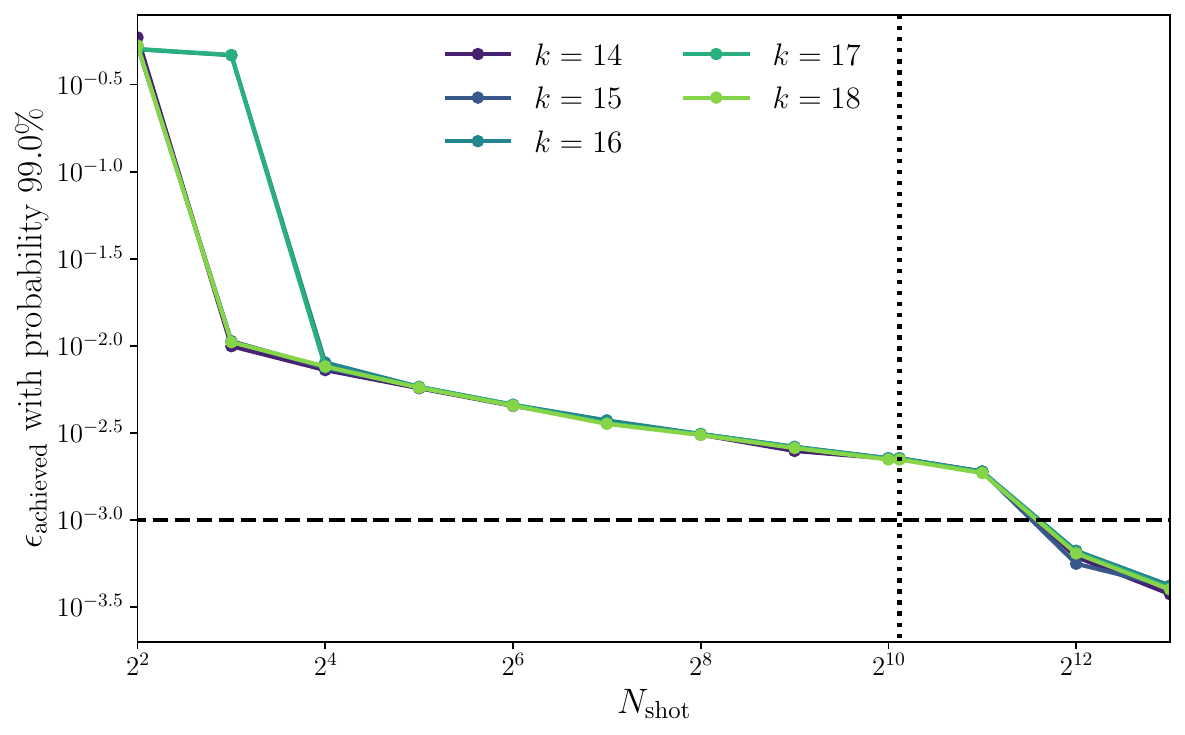}\label{fig:exceptionalvalues16}}
\subfigure[]{\includegraphics[width=0.49\columnwidth]{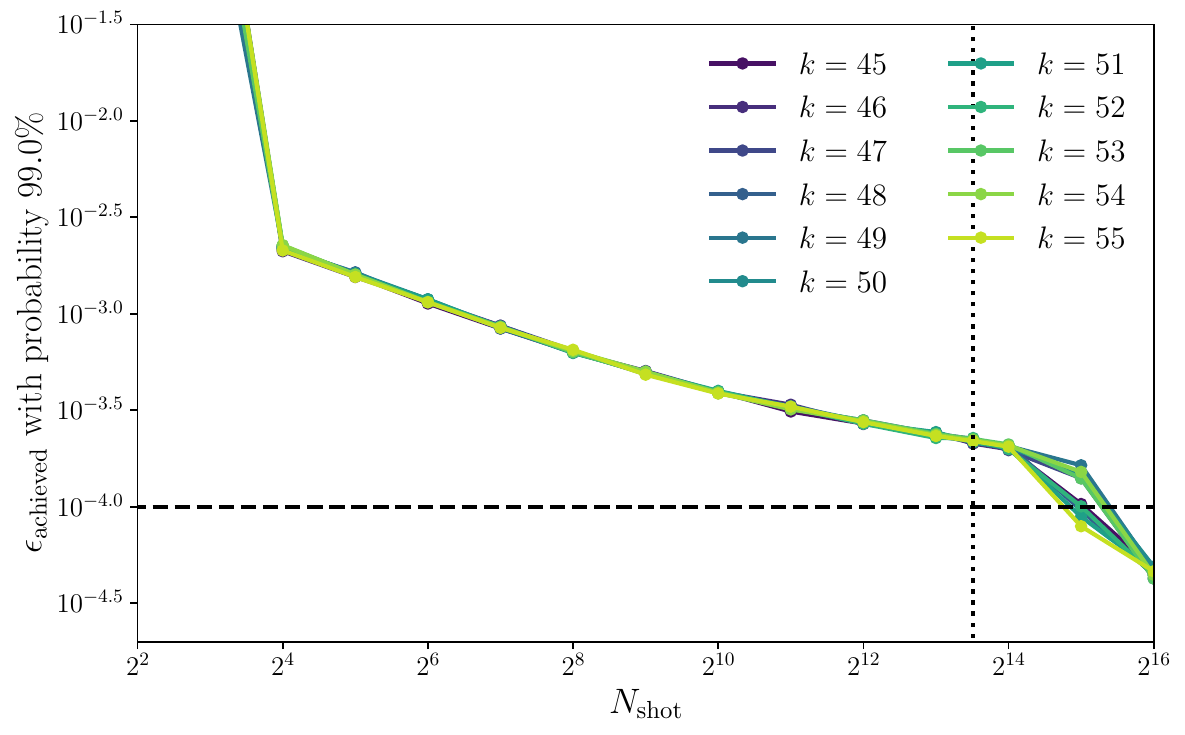}\label{fig:exceptionalvalues50}}
\caption{\label{fig:normalexceptionalvalues} 
Each data point represents the minimum precision $\epsilon_\mathrm{achieved}$ achieved with probability at least $1-\delta$, estimated from $10\,000$ runs of MLQAE, for various values of the number of shots $N_\mathrm{shot}$ and for various values of the amplitude $a$.
The $N_\mathrm{shot}$ values used are powers of 2, as well as an extra data point for the $N_\mathrm{shot}$ calculated according to  Eq. \eqref{eq:req_Nshot} for a particular target precision $\epsilon$, which are also indicated by the vertical dotted and horizontal dashed lines respectively.
The left column ((a) and (c)) uses the parameters $d=16$, $\epsilon=10^{-3}$, $\delta=0.01$, while the right column ((b) and (d)) uses the parameters $d=50$, $\epsilon=10^{-4}$, $\delta=0.01$.
The top row ((a) and (b)) shows \textit{typical} values of the amplitude $a$, chosen as $a=0.1,0.2\dots,0.8, 0.9$ as well as the results for choosing $a$ randomly from the full range for each of the $10\,000$ runs.
The bottom row ((c) and (d)) shows \textit{exceptional} values of the amplitude $a$, chosen as $a = a^{(d\rightarrow p_d(\theta_a)=0,1)}_k+\epsilon$ see [Eq. \eqref{eq:exceptional_values}], with (c) $k=14,15,16,17,18$ and (d) $k=45,46,\dots,54,55$ in order to be reasonably close to $a=0.5$.
For each plot, we performed the numerical likelihood maximisation with a grid of $3/\epsilon_\mathrm{min}$ equally-spaced points, where $\epsilon_\mathrm{min}$ is the precision expected for the largest value of $N_\mathrm{shot}$ plotted.
}
\end{centering}
\end{figure}

In Fig. \ref{fig:normalexceptionalvalues}, each data point represents the minimum precision $\epsilon_\mathrm{achieved}$ achieved with probability at least $1-\delta$, estimated from $10\,000$ runs of MLQAE, for various values of the number of shots $N_\mathrm{shot}$ and for various values of the amplitude $a$.
The $N_\mathrm{shot}$ values used are powers of 2, as well as an extra data point for the $N_\mathrm{shot}$ calculated according to Eq. \eqref{eq:req_Nshot} for a particular target precision $\epsilon$, which are also indicated in each plot by the vertical dotted and horizontal dashed lines respectively.
The left column uses the parameters $d=16$, $\epsilon=10^{-3}$, $\delta=0.01$, while the right column uses the parameters $d=50$, $\epsilon=10^{-4}$, $\delta=0.01$.
The top (bottom) row shows some \textit{typical} (\textit{exceptional}) values of the amplitude $a$.
For each plot in Fig. \ref{fig:normalexceptionalvalues}, we performed the numerical likelihood maximisation with a grid of $3/\epsilon_\mathrm{min}$ equally-spaced points, where $\epsilon_\mathrm{min}$ is the precision expected for the largest value of $N_\mathrm{shot}$ plotted.
It can be seen that for the typical values of the amplitude $a$, the target precision $\epsilon$ is achieved at values of $N_\mathrm{shot}$ that are lower than or close to the value predicted by Eq. \eqref{eq:req_Nshot}.
For the exceptional values of the amplitude $a$, however, it can be seen that values of $N_\mathrm{shot}$ that are between $2$ and $4$ times larger than the value predicted by Eq. \eqref{eq:req_Nshot} are required to achieve the target precision $\epsilon$.
This is consistent with the plots in Fig. \ref{fig:scatter}, and the reasons for these exceptional values will be explored in the next section.

A similar observation was made in \citep{tanaka2021amplitude} in the case of MLQAE in a noisy setting, where a multiparameter likelihood maximisation is performed that includes the noise rate as a nuissance parameter.
In that work, `anomalous target' values of the amplitude $a$ are observed for which the algorithm fails to achieve the desired precision; however, those anomalous target values are identified as being caused by a failure of the multiparameter likelihood maximisation procedure to accurately estimate the noise rate, which is not relevant in the noiseless case where the likelihood-maximisation is performed only in one dimension.
Thus, the exceptional values seen here are a distinct phenomenon to the anomalous target values seen in \citep{tanaka2021amplitude}.

\section{Exceptional values}
\label{sec:exceptionalvalues}

In subsection \ref{ssec:numval}, we showed numerically that choosing the number of shots $N_\mathrm{shot}$ according to Eq. \eqref{eq:req_Nshot} approximately achieves the target precision $\epsilon$ with probability at least $1-\delta$ for \textit{typical} values of the amplitude $a$, but that it fails for particular \textit{exceptional} values in the vicinity of $a=a^{(d\rightarrow p_d(\theta_a)=0,1)}_k$ for which the maximum Grover-depth $d$ produces `good' states with probability $1$ or $0$.
To see why this problem occurs, it is helpful to examine the likelihood-maximisation process at a conceptual level.

\begin{figure}
\begin{centering}
\subfigure[]{\includegraphics[width=0.49\columnwidth]{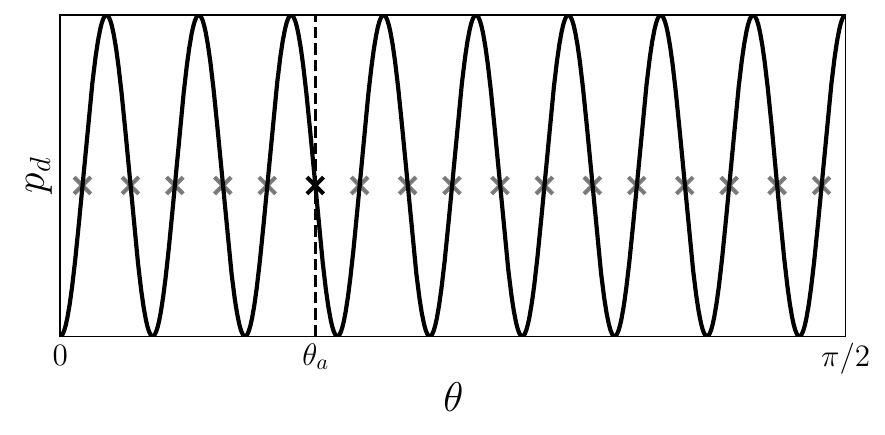}\label{fig:exceptionalvaluesexplanationprobs_typical}}
\subfigure[]{\includegraphics[width=0.49\columnwidth]{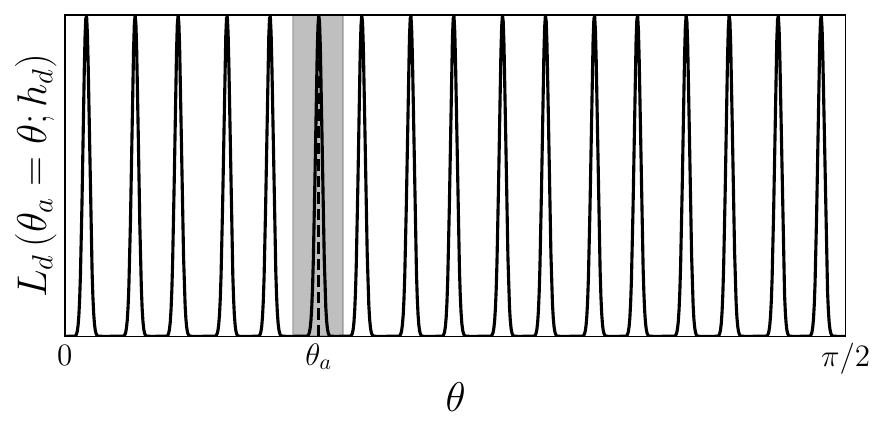}\label{fig:exceptionalvaluesexplanationllf_typical}}
\subfigure[]{\includegraphics[width=0.49\columnwidth]{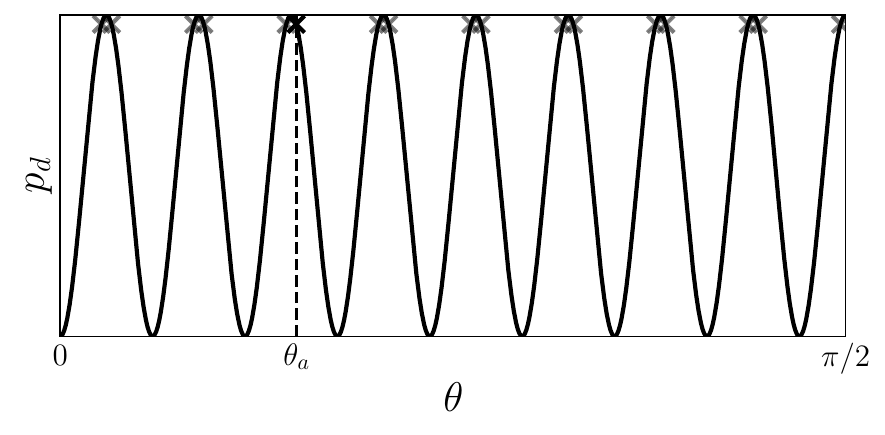}\label{fig:exceptionalvaluesexplanationprobs_exceptional}}
\subfigure[]{\includegraphics[width=0.49\columnwidth]{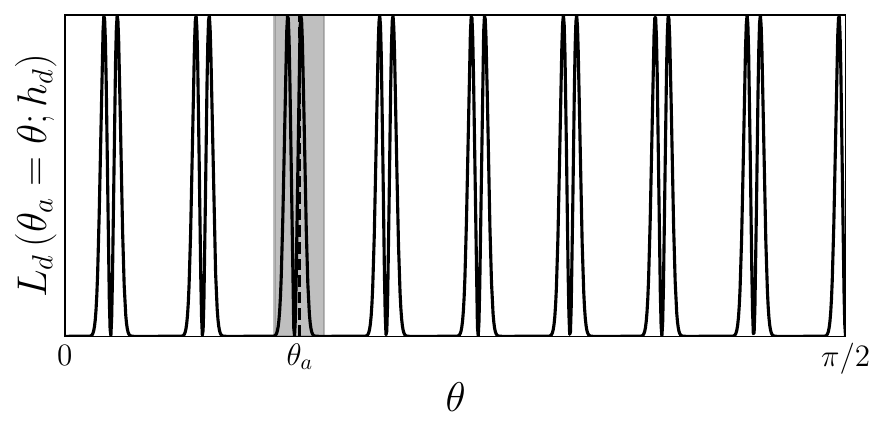}\label{fig:exceptionalvaluesexplanationllf_exceptional}}
\caption{\label{fig:exceptionalvaluesexplanation} 
A qualitative illustration of the emergence of exceptional values of the amplitude $a$ in the vicinity of $a=a^{(d\rightarrow p_d(\theta_a)=0,1)}_k$ [see Eq. \eqref{eq:exceptional_values}].
The top row illustrates the case of typical values of the amplitude $a$ (and of the corresponding angle $\theta_a$). Plot (a) illustrates the probability $p_d$ of measuring a `good' state after running the circuit with the maximum Grover-depth, $d$.
The true value of the angle $\theta_a$ is indicated by the black cross and the vertical dashed line, while other angles that produce the same probability are indicated by grey crosses.
As the typical values occur far from $p_d=0$ and $p_d=1$, the corresponding Grover-depth $d$ likelihood function $L_d$ typically looks as in plot (b), where the likelihood peaks correspond to the crosses from plot (a).
The Grover-depth $d$ measurement data does not contain the information to determine which of the peaks correspond to the true angle $\theta_a$, but the information gained from the lower Grover-depth circuits is typically sufficient to narrow $\theta$ down to a small range, indicated by the shaded area in plot (b), that contains only one of these peaks.
The bottom row ((c) and (d)) illustrates the case of exceptional values, in this example close to $p_d=1$.
As some pairs of crosses occur close together in plot (c), some pairs of peaks occur close together in the corresponding likelihood function in plot (d), including two peaks that both appear in the shaded region, meaning that it is not possible to determine which of the two peaks correspond to the true angle $\theta_a$.
}
\end{centering}
\end{figure}

To this end, Fig. \ref{fig:exceptionalvaluesexplanation} shows a qualitative illustration of the emergence of exceptional in the vicinity of $a=a^{(d\rightarrow p_d(\theta_a)=0,1)}_k$ [see Eq. \eqref{eq:exceptional_values}].
The top row illustrates the case of typical values of the amplitude $a$ (and of the corresponding angle $\theta_a$).
Fig. \ref{fig:exceptionalvaluesexplanationprobs_typical} illustrates the probability $p_d(\theta_a)$ of measuring a `good' state after running the circuit with the maximum Grover-depth, $d$.
The true value of the angle $\theta_a$ is indicated by the black cross and the vertical dashed line, while other angles that produce the same probability are indicated by grey crosses.
As the typical values occur far from $p_d(\theta_a)=0$ and $p_d(\theta_a)=1$, the corresponding Grover-depth $d$ likelihood function $L_d$ typically looks as in Fig. \ref{fig:exceptionalvaluesexplanationllf_typical}, where the likelihood peaks correspond to the crosses from plot (a).
The Grover-depth $d$ measurement data does not contain the information to determine which of the peaks correspond to the true angle $\theta_a$, but the information gained from the lower Grover-depth circuits is typically sufficient to narrow $\theta$ down to a small range, indicated by the shaded area in plot (b), that contains only one of these peaks.

\begin{figure}
\begin{centering}
\includegraphics[width=0.75\columnwidth]{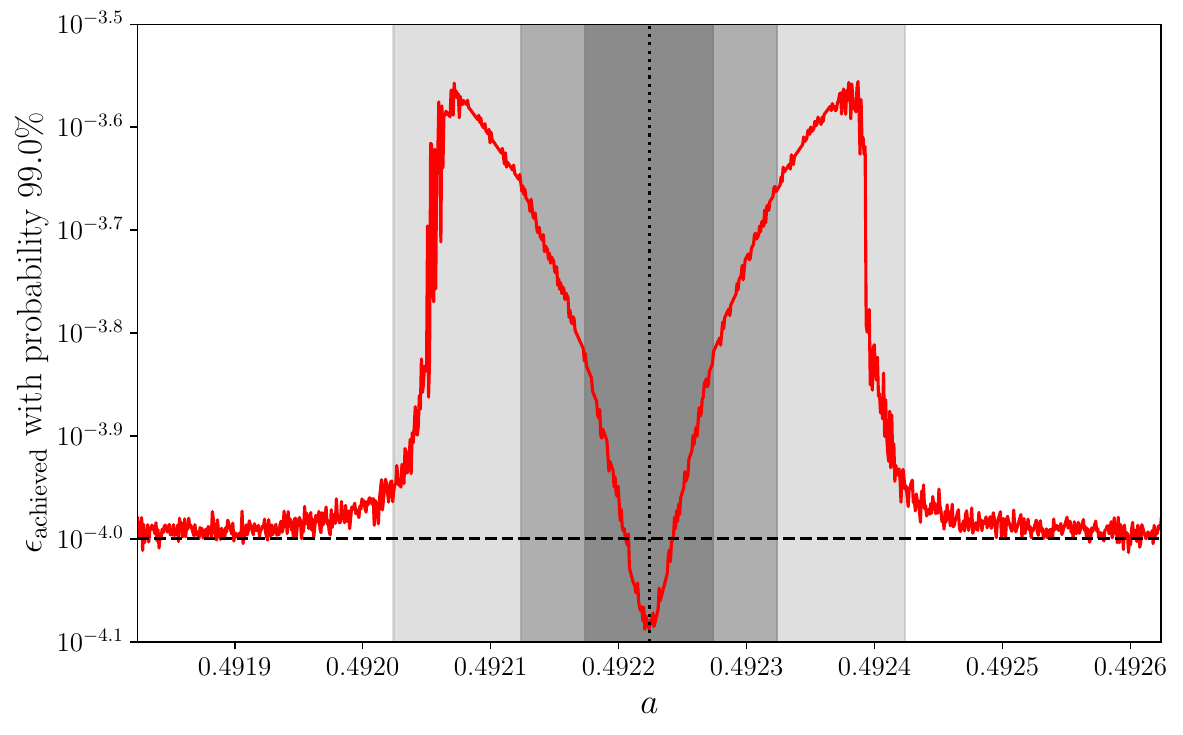}
\caption{\label{fig:exceptionalregion50} 
The red line shows the precision $\epsilon_\mathrm{achieved}$ achieved with probability $1-\delta=99\%$ (estimated from $10\,000$ MLQAE runs), when targeting a precision $\epsilon=10^{-4}$, indicated by the horizontal dashed line, using an exponential Grover-depth schedule with maximum Grover-depth $d=50$, at $1000$ different values of the amplitude $a$ in the vicinity of $a^{(d\rightarrow p_d(\theta_a)=0,1)}_k$ with $k=50$, indicated by the vertical dotted line, and the shaded regions are centred on this value.
The numerical likelihood maximisation has been performed by calculating the likelihood function at $\sim30/\epsilon$ equally spaced values of $\theta$, a much finer search than is needed in practice in order to reduce the size of the fluctuations in the plot.
The darkest shaded region in the centre has width $\epsilon$, while the medium and lightest shaded regions have width $2\epsilon$ and $4\epsilon$ respectively.
}
\end{centering}
\end{figure}

However, the bottom row illustrates the case of an exceptional value, which in this example is close to $p_d(\theta_a)=1$.
As some pairs of crosses occur close together in Fig. \ref{fig:exceptionalvaluesexplanationprobs_exceptional}, some pairs of peaks occur close together in the corresponding likelihood function in Fig. \ref{fig:exceptionalvaluesexplanationllf_exceptional}, including two peaks that both appear in the shaded region, meaning that it is not possible to determine which of the two peaks correspond to the true angle $\theta_a$.
If the distance $|\tilde{a} - \tilde{a}_\mathrm{False}|$ between the correct estimate $\tilde{a}$ and the nearby `false' estimate $\tilde{a}_\mathrm{False}$ is much smaller than the target precision, that is if $|\tilde{a} - \tilde{a}_\mathrm{False}| \ll \epsilon$, then either estimate is likely to work well, and this ambiguity will not cause an issue.
In the opposite extreme, if $|\tilde{a} - \tilde{a}_\mathrm{False}| \gtrsim 2\epsilon$, then the false estimate $\tilde{a}_\mathrm{False}$ is unlikely to be in the ambiguous region (illustrated by the shaded regions in Fig. \ref{fig:exceptionalvaluesexplanation}), as $2\epsilon$ is approximately the precision that would be achieved if the data from the maximum Grover-depth $d$ runs were not included in the analysis.
The exceptional values are thus expected to occur roughly in the intermediate regime, where $\epsilon\lesssim |\tilde{a} - \tilde{a}_\mathrm{False}| \lesssim 2\epsilon$.
In Fig. \ref{fig:exceptionalregion50}, this corresponds to the expectation that the exceptional values should be approximately contained in the region of medium darkness.
This intuition is approximately confirmed, although the exceptional values do extend further out in to the lightest shaded region.

This can also be understood in terms of the Bernstein-von Mises theorem.
The starting point of the derivation of the required number of shots $N_\mathrm{shot}$ in Eq. \eqref{eq:req_Nshot} was the assumption that the regularity conditions set out in the Bernstein-von Mises theorem hold.
The theorem is neatly explained and summarised by Giurgica-Tiron et al. in \citep{giurgicatiron2022low}, and the conditions can be expressed most simply as the following three requirements:
\begin{enumerate}
    \item \label{bvm_lf_smooth} The likelihood function $L(\theta_a=\theta;\boldsymbol{h})$ and its logarithm should be smooth over the relevant interval of $\theta$.
    \item \label{bvm_i_smooth} The Fisher information $I(a)$ should be smooth over the relevant interval of $a$ (and thus of $\theta$).
    This implies that that the logarithm of the likelihood function should be at least thrice differentiable
    \item \label{bvm_prior} The first derivative of the prior distribution should exist and be continuous.
\end{enumerate}
In the present case, the prior distribution is uniform, and so condition \ref{bvm_prior} is trivially satisfied.
However, it can be seen from Eq. \eqref{eq:singledepth_lf}, Eq. \eqref{eq:lf} and Fig. \ref{fig:mlqae_likelihood} that the likelihood function $L(\theta_a=\theta;\boldsymbol{h})$ goes to $0$ at the amplitude values $a^{(d_j\rightarrow P_\mathrm{good}=0,1)}_k$ if the measurement record $h_{d_j}$ contains both `good' and `bad' outcomes, meaning that at these points the logarithm of the likelihood function diverges to negative infinity, violating conditions \ref{bvm_lf_smooth} and \ref{bvm_i_smooth}.
If the information gained from measurements at lower Grover-depths is sufficient to restrict the likely estimates $\tilde{a}$ to an interval around the amplitude $a$ that does not include any $a^{(d_j\rightarrow p_d(\theta_a)=0,1)}_k$ values, the validity of conditions \ref{bvm_lf_smooth} and \ref{bvm_i_smooth} is recovered in this interval, but if this is not the case than the Bernstein-von Mises theorem does not apply, explaining why using a number of shots $N_\mathrm{shot}$ according to Eq. \eqref{eq:req_Nshot} is not even approximately sufficient for the exceptional values.
From this perspective, exceptional values should occur near the $a^{(d_j\rightarrow p_d(\theta_a)=0,1)}_k$ at any Grover-depth $d_j$ in the depth schedule, but in practice those at depths smaller than the maximum depth $d$ have a much less significant impact due to the lower information content of measurements at those depths.

\section{Depth jittering heuristic}
\label{sec:depthjitter}

As discussed in section \ref{sec:exceptionalvalues}, exactly which values of the amplitude value $a$ are exceptional depends on exactly which Grover-depths appear in the Grover-depth schedule $D$, with a particular importance associated with the maximum depth $d$.
With this in mind, we propose a heuristic method to produce a modified Grover-depth schedule $D'$ in which the $N_\mathrm{shot}$ shots for some Grover-depths $d_j \in D_{\mathrm{EXP},\nu}$ are instead spread out over a small number of nearby depths, in order to mitigate the effects of exceptional values.
We refer to this process as `depth jittering', and the jittered Grover-depth schedule $D'$ can be produced from the original Grover-depth schedule $D$ via Algorithm \ref{alg:depthjitter}.

\begin{figure}[htb]
    \centering
    \begin{minipage}{0.89\textwidth}
        \begin{algorithm}[H]
                \caption{Depth jittering algorithm\label{alg:depthjitter}}
                \textbf{Inputs}: $D$: Original Grover-depth schedule, with size at least 2, in ascending order. \newline 
                \hspace*{38pt} $c$: Logarithmic spread coefficient. \newline
                \textbf{Outputs}: $D'$: Jittered Grover-depth schedule, in ascending order. \newline
                \hspace*{\algorithmicindent}\hspace*{31pt} $F^{(\mathrm{shot})}$: Corresponding shot fractions.
                \begin{algorithmic}[1]
                \State \textbf{initialise} $D' \coloneqq \{\}$, $F^{(\mathrm{shot})} \coloneqq \{\}$
                \State \textbf{for} $d_j$ in $\mathtt{reverse}(D)$: \textcolor{gray}{\texttt{\#reverse order to prioritise jittering larger depths}}
                \State \hspace{20pt} \textbf{initialise} $\mathtt{jitter}\coloneqq\mathtt{False}$
                \State \hspace{20pt} calculate  $\mathtt{depth\_spread}\coloneqq\mathrm{Round}\left[\ln\left(c d_j\right)\right]$
                \State \hspace{20pt} \textbf{if} $d_j=\mathtt{max}(D)$:
                \State \hspace{40pt} \textbf{set} $\mathtt{lower}\coloneqq d_j - \mathtt{depth\_spread}$
                \State \hspace{40pt} \textbf{set} $\mathtt{upper}\coloneqq d_j$ \textcolor{gray}{\texttt{\#don't go beyond maximum depth}}
                \State \hspace{40pt} \textbf{if} $\mathtt{lower} > d_{j-1} +1$: \textbf{set} $\mathtt{jitter}\coloneqq\mathtt{True}$
                \State \hspace{20pt} \textbf{elif} $\mathtt{min}(D) < d_j < \mathtt{max}(D)$:
                \State \hspace{40pt} \textbf{set} $\mathtt{lower}\coloneqq d_j - \mathtt{depth\_spread}$ and $\mathtt{upper}\coloneqq d_j + \mathtt{depth\_spread}$
                \State \hspace{40pt} \textbf{if} $\mathtt{lower} > d_{j-1} +1$ \textbf{and} $\mathtt{upper} < \mathtt{min}(D')-1$: \textbf{set} $\mathtt{jitter}\coloneqq\mathtt{True}$
                \State \hspace{20pt} \textbf{elif} $d_j=\mathtt{min}(D)$ \textbf{and not} $d_j=0$: \textcolor{gray}{\texttt{\#jittering at depth 0 not necessary}}
                \State \hspace{40pt} \textbf{set} $\mathtt{lower}\coloneqq \mathtt{max}(0, d_j - \mathtt{depth\_spread})$ \textcolor{gray}{\texttt{\#don't go below 0 depth}}
                \State \hspace{40pt} \textbf{set} $\mathtt{upper}\coloneqq d_j + \mathtt{depth\_spread}$
                \State \hspace{40pt} \textbf{if} $\mathtt{upper} < \mathtt{min}(D')-1$: \textbf{set} $\mathtt{jitter}\coloneqq\mathtt{True}$
                \State \textbf{if} \texttt{jitter:}
                \State \hspace{20pt} \textbf{for} $(d_j)'_k$ from \texttt{upper} to \texttt{lower}: \textcolor{gray}{\texttt{\#add in descending order to be reversed at end}}
                \State \hspace{40pt} \textbf{add}  $(d_j)'_k$ to $D'$ and $1/\left(\mathtt{upper}-\mathtt{lower} + 1\right)$ to $F^{(\mathrm{shot})}$
                \State \textbf{else}: 
                \State \hspace{20pt} \textbf{add} $d_j$ to $D'$ and $1$ to $F^{(\mathrm{shot})}$
                \State \textbf{set} $D'\coloneqq\mathtt{reverse}(D')$ and  $F^{(\mathrm{shot})}\coloneqq\mathtt{reverse}(F^{(\mathrm{shot})})$
                \State \textbf{return} $D'$ and $F_\mathrm{shot}$
                \end{algorithmic}
        \end{algorithm}
    \end{minipage}
\end{figure}

In short, Algorithm \ref{alg:depthjitter} takes each Grover-depth $d_j$ in the original schedule $D$, as well as a `logarithmic spread coefficient' $c$ and adds all the depths between $\sim \left(d_j-\log\left(c d_j\right)\right)$ and $\sim \left(d_j+\log\left(c d_j\right)\right)$ to a new `jittered' schedule $D'$.
This is not done for zero-depth $d_j=0$, as this depth has no exceptional values.
This is also not done for depths $d_j$ where this spread would overlap with the spreads from neighbouring depths $d_{j-1}$ and $d_{j+1}$, where jittering the larger depths is prioritised in these cases.
For the maximum Grover-depth $d$, only the part of the spread that is less than or equal to $d$ is included, to avoid exceeding the maximum Grover-depth, while for the minimum Grover-depth the spread is truncated if necessary to avoid impossible depths below 0.

The use of depth-spreads of logarithmic size is a heuristic choice, motivated by the need to preserve the approximate scaling results presented in previous sections.
Algorithm \ref{alg:depthjitter} also outputs a set of `shot fractions' $F^{(\mathrm{shot})}$ that simply keeps track of the factor by which $N_\mathrm{shot}$ should be scaled down at each new of the new depths $d'_j \in D'$ in order to effectively spread the $N_\mathrm{shot}$ shots that would have originally been run at depth $d_j$ over the relevant jittered depths.
Taking the examples of exponential schedules $D_{\mathrm{EXP},\nu}$ with maximum depths $d=16$ and $d=50$ (as in Fig. \ref{fig:scatter} and Fig. \ref{fig:normalexceptionalvalues}) and setting the spread coefficient $c=2$ (this choice is arbitrary, but seems to work well in practice), the original schedules are 
\begin{align}
    D_{\mathrm{EXP},\nu} & = \{0, 1, 2, 4, 8, 16\} \\
    D_{\mathrm{EXP},\nu} & = \{0, 1, 2, 4, 7, 14, 26, 50\},
\end{align}
while the jittered schedules are
\begin{align}
    D'_{\mathrm{EXP},\nu} & = \{0; 1; 2; 4; 8; 13, 14, 15, 16\} \\
    D'_{\mathrm{EXP},\nu} &= \{0; 1; 2; 4; 7; 11, 12, 13, 14, 15, 16, 17; \nonumber\\
    &{\hspace{17pt}} 22, 23, 24, 25, 26, 27, 28, 29, 30; 45, 46, 47, 48, 49, 50\},
\end{align}
where semicolons have been used to separate the jittered depths into groups corresponding to the original depths.

It is also necessary to define jittered versions of the quantities $S_D^{(1)}$ and $S_D^{(2)}$ from Eq. \eqref{eq:SD1} and Eq. \eqref{eq:SD1}.
These are simply
\begin{align}
     S_{D'}^{(1)'} &\equiv \sum_{j=0}^{q-1} F^{(\mathrm{shot})}_j(2d_j' + 1)
     \label{eq:SD1jitter}
\end{align}
and
\begin{align}
     S_{D'}^{(2)'} &\equiv \sqrt{\sum_{j=0}^{q-1}F^{(\mathrm{shot})}_j(2d_j' + 1)^2}.
     \label{eq:SD2jitter}
\end{align}
With these definitions, the jittered number of shots $N_\mathrm{shot}'$ can be calculated according to Eq. \eqref{eq:req_Nshot}, which is in general slightly different from the original $N_\mathrm{shot}$.
When the jittered MLQAE algorithm is run, the actual number of shots used at depth $d_j'$ is $\lceil F^{(\mathrm{shot})}_jN'_\mathrm{shot} \rceil$.

\begin{figure}
\begin{centering}
\includegraphics[width=0.75\columnwidth]{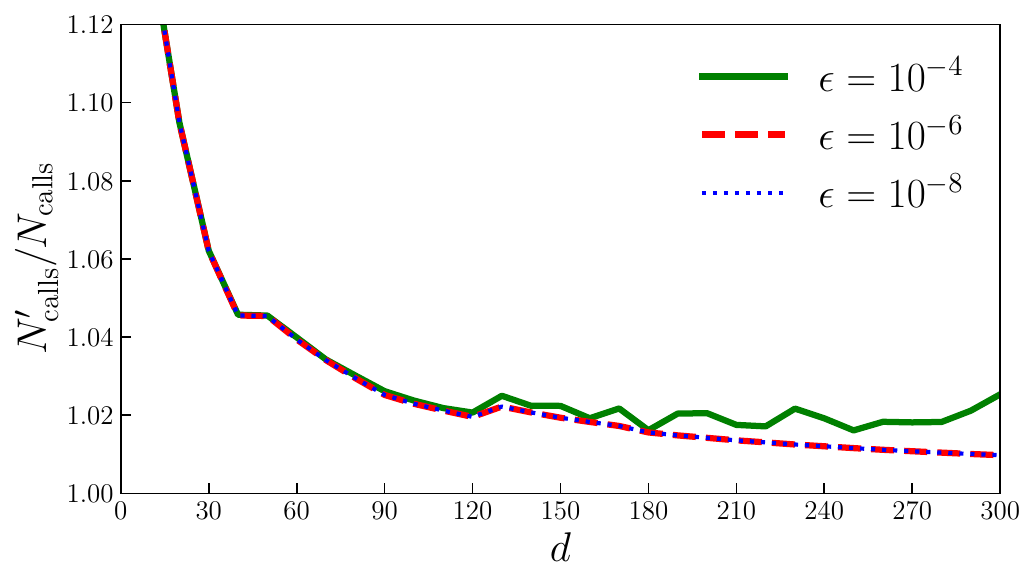}
\caption{\label{fig:jitterunjittercallratio} 
The ratio of the number of calls $N_\mathrm{calls}'$ and $N_\mathrm{calls}$ to the algorithm $\hat{A}$ in the case of a jittered and unjittered Grover-depth schedule respectively as a function of maximum Grover-depth $d$, for three different choices of the target precision $\epsilon$. An acceptable failure probability of $\delta=0.01$ is used in all cases.
The visible increase at the highest depths for the $\epsilon=10^{-4}$ case (solid green line) occurs because of the ceiling function applied in the expression $\lceil F^{(\mathrm{shot})}_jN'_\mathrm{shot} \rceil$ for the actual number of shots used at depth $d_j'$, as $N'_\mathrm{shot}$ becomes small.
}
\end{centering}
\end{figure}

To demonstrate that the extent of the speedup for MLQAE compared to classically sampling from the outputs of the algorithm $\hat{A}$ is not significantly affected by the depth-jittering procedure, Fig. \ref{fig:jitterunjittercallratio} shows the ratio of the number of calls $N_\mathrm{calls}'$ and $N_\mathrm{calls}$ to the algorithm $\hat{A}$ in the case of a jittered and unjittered Grover-depth schedule respectively as a function of maximum Grover-depth $d$, for three different choices of the target precision $\epsilon$. An acceptable failure probability of $\delta=0.01$ is used in all cases.
The visible increase at the highest depths for the $\epsilon=10^{-4}$ case (solid green line) occurs because of the ceiling function applied in the expression $\lceil F^{(\mathrm{shot})}_jN'_\mathrm{shot} \rceil$ for the actual number of shots used at depth $d_j'$, as $N_\mathrm{shot}$ becomes small.
For the more strict precision requirements (the dashed red and dotted blue lines) where $N_\mathrm{shot}$ is not small at the larger Grover-depths, the ratio is convincingly tending toward a value of $1$.

\subsection{Numerical validation}
\label{ssec:numvaljitter}
\begin{figure}
\begin{centering}
\subfigure[]{\includegraphics[width=0.49\columnwidth]{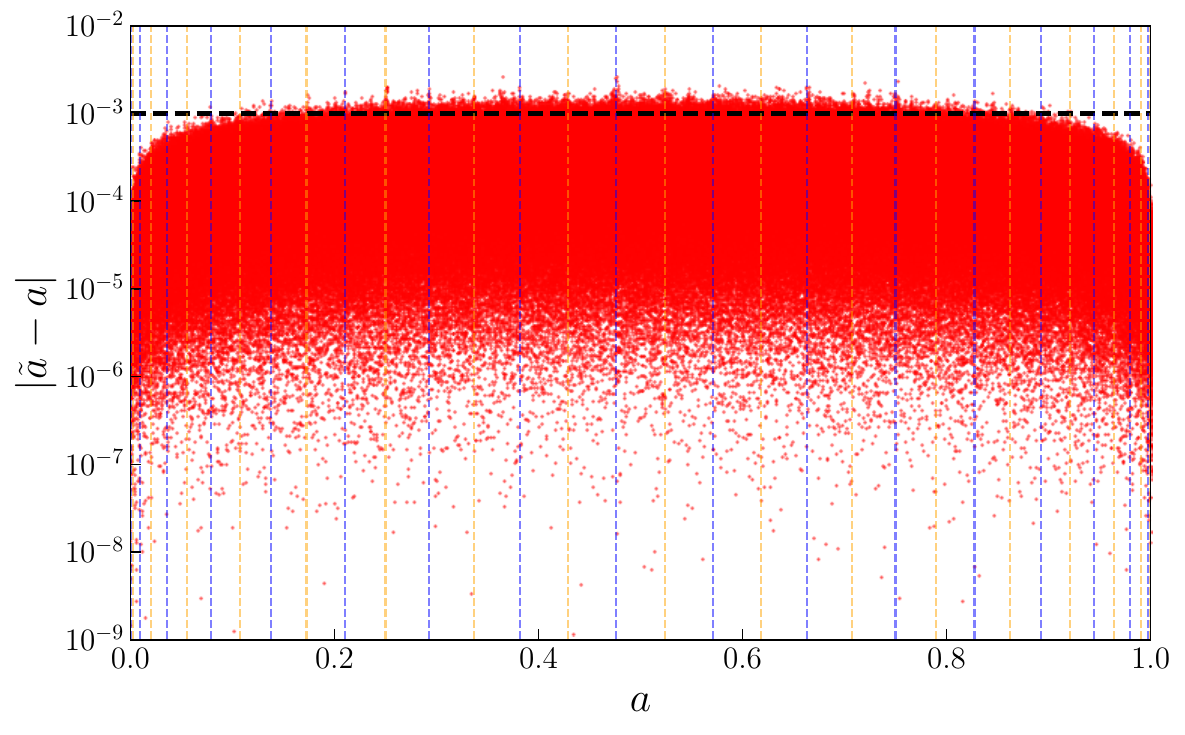}\label{fig:scatter16jitter}}
\subfigure[]{\includegraphics[width=0.49\columnwidth]{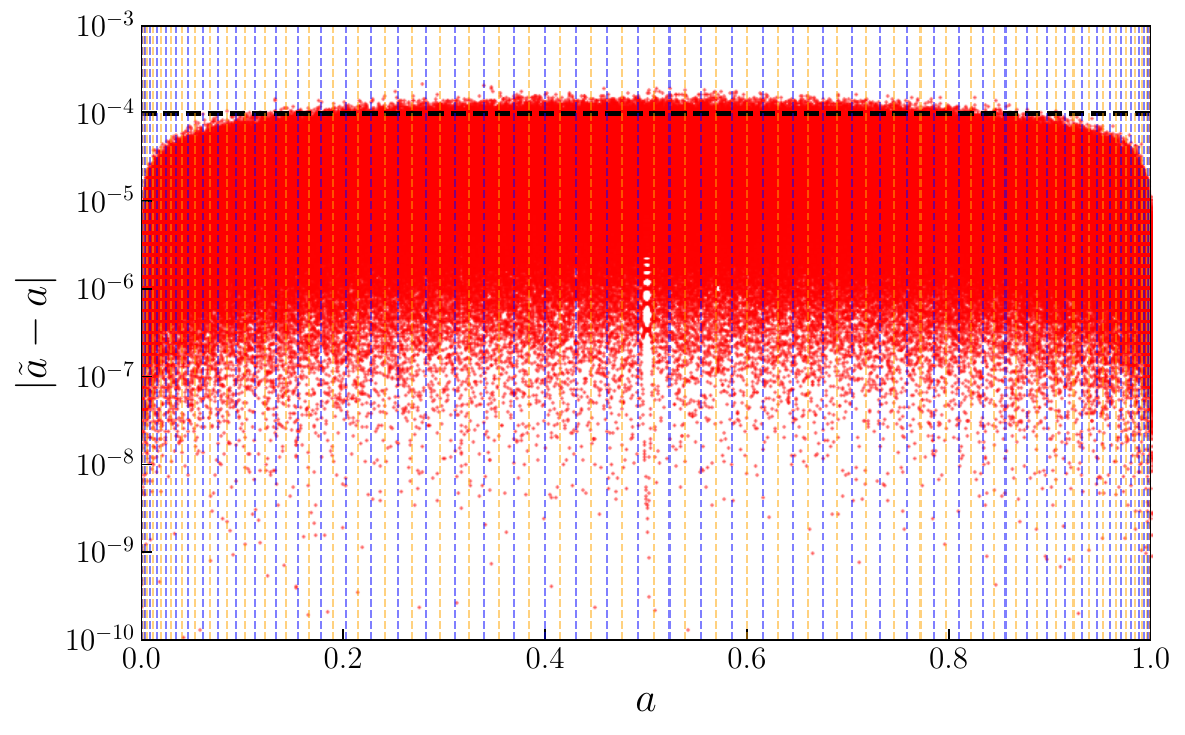}\label{fig:scatter50jitter}}
\caption{\label{fig:scatterjitter} 
Data produced by exactly the same procedure and with the same parameters as in Fig. \ref{fig:scatter}, but with a jittered Grover-depth schedule $D_{\mathrm{EXP},\nu}'$.
The sharp features in Fig. \ref{fig:scatter} associated with exceptional values of the amplitude $a$ have been largely removed by the depth-jittering procedure.
}
\end{centering}
\end{figure}

\begin{figure}
\begin{centering}
\subfigure[]{\includegraphics[width=0.49\columnwidth]{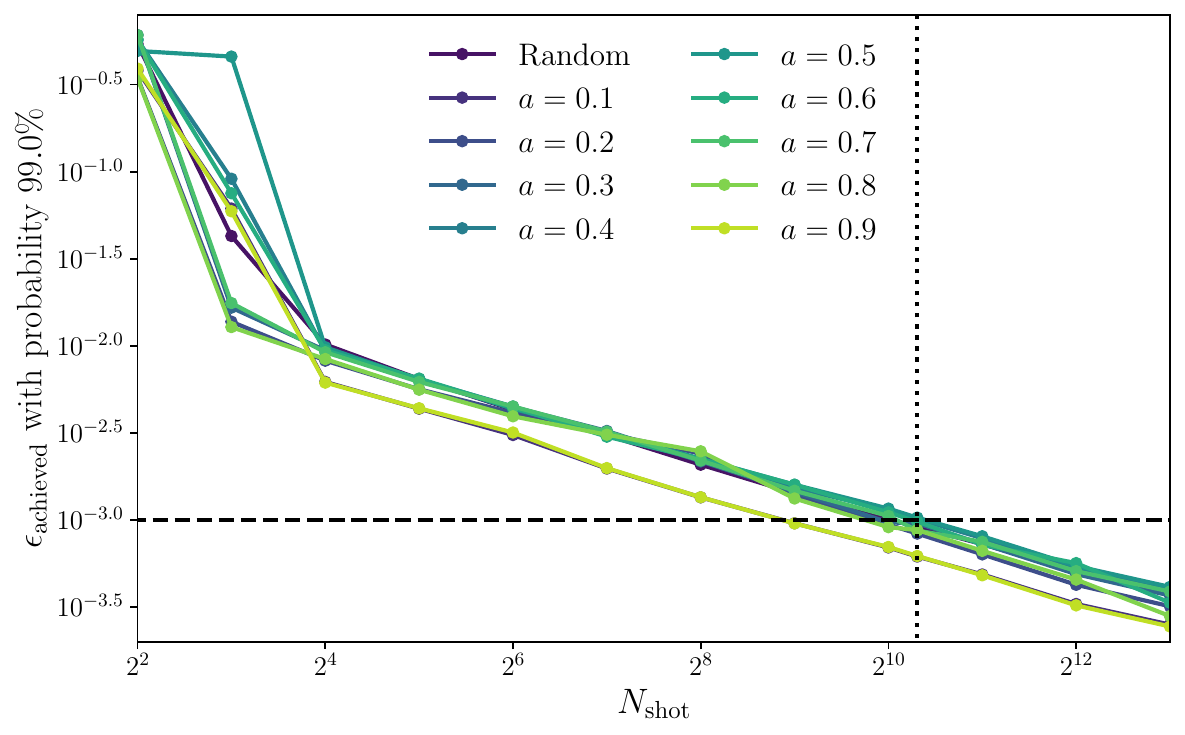}\label{fig:normalvalues16jitter}}
\subfigure[]{\includegraphics[width=0.49\columnwidth]{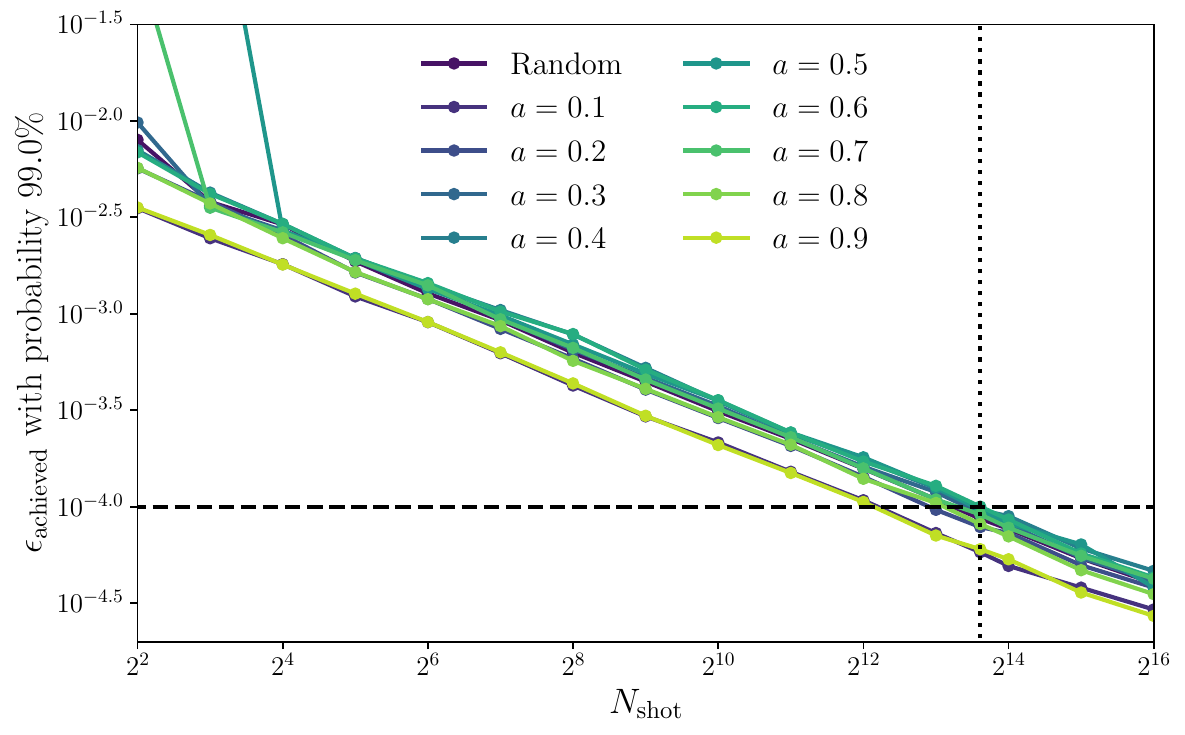}\label{fig:normalvalues50jitter}}
\subfigure[]{\includegraphics[width=0.49\columnwidth]{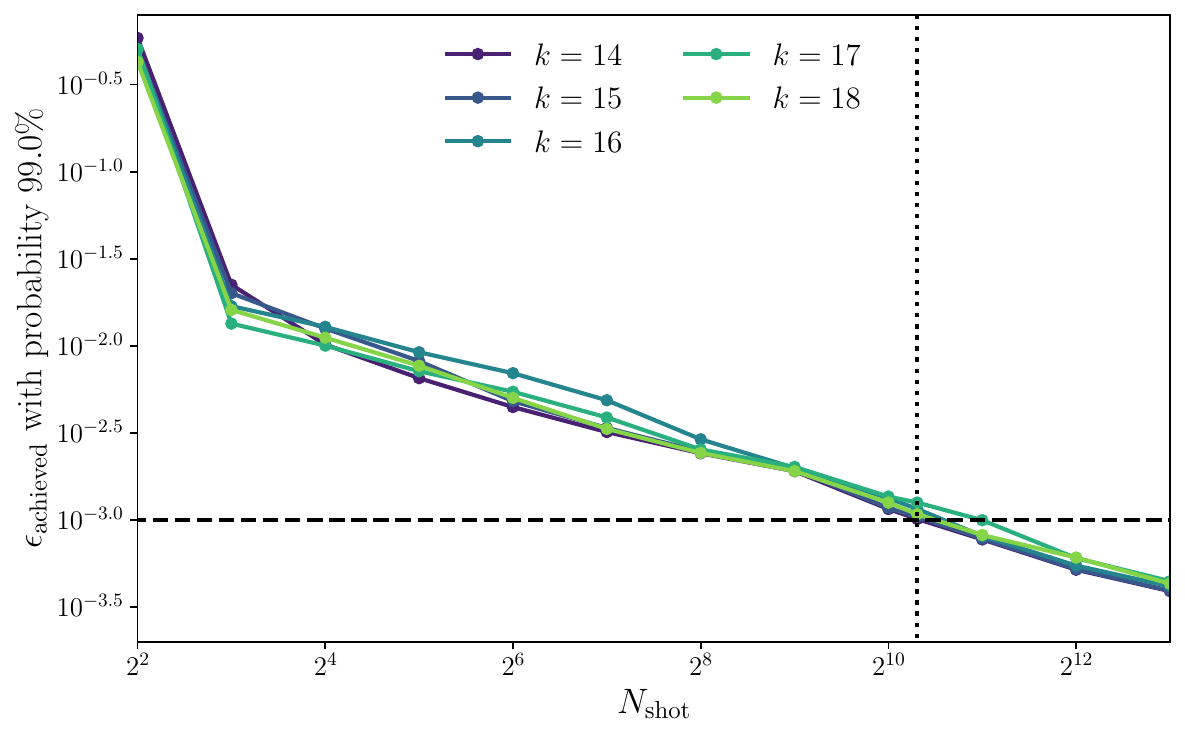}\label{fig:exceptionalvalues16jitter}}
\subfigure[]{\includegraphics[width=0.49\columnwidth]{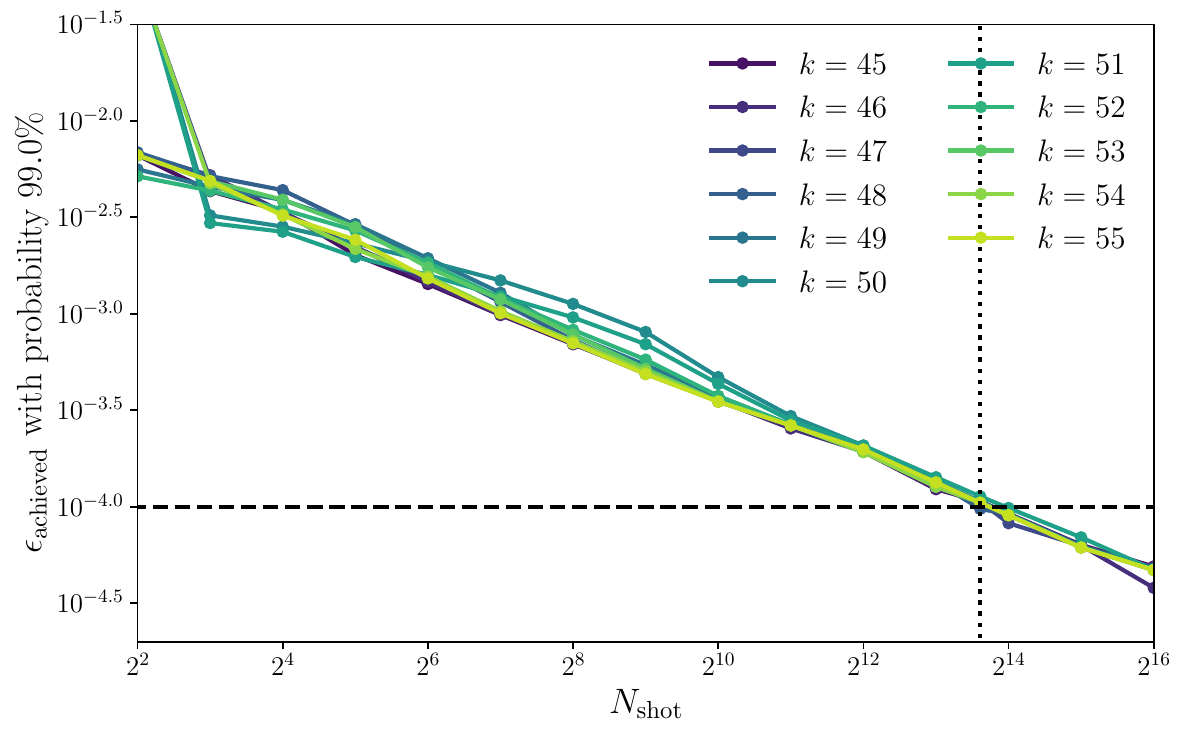}\label{fig:exceptionalvalues50jitter}}
\caption{\label{fig:normalexceptionalvaluesjitter} 
Data produced by exactly the same procedure and with the same parameters as in Fig. \ref{fig:normalexceptionalvalues}, but with a jittered Grover-depth schedule $D_{\mathrm{EXP},\nu}'$.
The MLQAE algorithm now achieves the target precision $\epsilon$ at an $N_\mathrm{shot}'$ value much closer to the value predicted by Eq. \eqref{eq:req_Nshot} for the exceptional values of the amplitude $a$ [(c) and (d)], without significantly affecting the performance for the typical values of the amplitude $a$ [(a) and (b)].
}
\end{centering}
\end{figure}

To show that the depth-jittering heuristic is effective, we have repeated the numerical experiments described in subsection \ref{ssec:numval} with exactly the same parameters, but using the depth-jittered schedule $D_{\mathrm{EXP},\nu}'$ instead of the original schedule $D_{\mathrm{EXP},\nu}$.
Fig. \ref{fig:scatterjitter} should be compared with Fig. \ref{fig:scatter}.
It can be seen that the sharp features associated with exceptional values of the amplitude $a$ have been largely removed by the depth-jittering procedure.
Likewise, Fig. \ref{fig:normalexceptionalvaluesjitter} should be compared with Fig. \ref{fig:normalexceptionalvalues}.
It can be seen that the MLQAE algorithm with the depth-jittering heuristic now achieves the target precision $\epsilon$ at an $N_\mathrm{shot}'$ value much closer to the value predicted by Eq. \eqref{eq:req_Nshot} for the exceptional values of the amplitude $a$ (Fig. \ref{fig:exceptionalvalues16jitter} and Fig. \ref{fig:exceptionalvalues50jitter}), without significantly affecting the performance for the typical values of the amplitude $a$ (Fig. \ref{fig:normalvalues16jitter} and Fig. \ref{fig:normalvalues50jitter}).

\section{Summary and further work}
\label{sec:summary}
In this paper, we have given new insights and made improvements to MLQAE, the maximum-likelihood quantum amplitude estimation algorithm first proposed by Suzuki et al. \citep{suzuki2020amplitude}.
Quantum amplitude estimation has the potential to be an important subroutine in quantum algorithms, offering quadratic speedups in many areas of science, engineering, and industry through its applications to quantum-enhanced Monte Carlo simulation \citep{montanaro2015quantum}, as well as to quantum machine learning \citep{kerenidis2019qmeans,wiebe2015quantum}.
The maximum-likelihood-based quantum amplitude estimation algorithm is one of a number of recent proposals that use much simpler quantum circuits than earlier work \citep{brassard2002quantum}, bringing quantum amplitude estimation closer to suitability for near- and mid-term quantum hardware.

The analysis in prior work focused on understanding how the average error scales with the number of calls to the quantum circuit whose output contains the amplitude we are trying to estimate, observing a quadratic speedup compared to simple classical sampling in the process.
In section \ref{sec:reqNshot}, we began our contribution by extending this analysis and showed how to calculate how many calls are required to target a particular precision with a particular choice of probability, rather than simply on average.
As with the prior work, our analysis assumed that the conditions required for the Bernstein-von Mises theorem to hold are satisfied, and was therefore not completely rigorous.
We also considered scenarios in which the circuit-depth cannot be made arbitrarily large, arguing that a trade-off between circuit-depth and quantum speedup developed in \citep{giurgicatiron2022low} can be achieved in a more a direct way.

While numerically validating our analysis in section \ref{sec:reqNshot} over the full range of target amplitudes, we observed that there are particular values of the amplitude,  \textit{exceptional values}, at which the algorithm fails to achieve the target precision.
In section \ref{sec:exceptionalvalues}, we described at which values of the amplitude these occur, and explained in detail why they occur, both formally in terms of violations of the Bernstein-von Mises theorem and via a more intuitive picture in terms of the likelihood maximisation process.

Having explained in section \ref{sec:exceptionalvalues} how the amplitudes which are the problematic exceptional values depended on the `Grover-depth' (the number of applications of a `Grover-like' operator present in the circuits that are run), we proposed in section \ref{sec:depthjitter} a heuristic method to mitigate the problem, which we refer to as `depth-jittering'.
This method involves taking the shots that would ordinarily all be performed on identical circuits at a single Grover-depth, and instead spreading them over a small number of nearby Grover-depths in order to bypass much of the sensitivity to Grover-depth.
We show numerically that this process has only a negligible effect on the algorithm runtime, and that the algorithm now achieves approximately the expected precision across the full range of target amplitudes and the  previously exceptional values are no longer problematic.

Our work leads naturally to a number of questions.
Most importantly, the only way in which we consider the effects of the noise that would be unavoidably present in near- and mid-term implementations of this algorithm is by imposing limitations on the available circuit depths, but prior work \citep{suzuki2020amplitude,brown2020quantum,tanaka2021amplitude,giurgicatiron2022low,tanaka2022noisy} has shown how to modify the likelihood functions to account for the noise on the quantum device.
It will be important to understand how these modifications would affect the exceptional values, as well as the effectiveness of our depth-jittering heuristic.
Another possible extension to our work is exploring the possibility of performing `adaptive' jittering, perhaps by detecting possible exceptional values through the low-depth measurements and only then jittering the high-depth runs if necessary.
The runs at each Grover-depth could still be performed in parallel, and so this advantage of MLQAE is not lost.
A further possible extension is to investigate whether there are numerical factors by which the number of measurement shots can be scaled up to reliably guard against the approximate nature of the theoretical analysis.
Overall, the insights we offer in this work help to improve the understanding of the maximum-likelihood quantum amplitude estimation algorithm, and move it toward being a useful tool for practical algorithms in the near- and mid-term.

\ack
AC thanks Oscar Higgott, George Umbrarescu, and Tom Dodd for helpful and interesting discussions.
The authors gratefully acknowledge support from the UKRI EPSRC Grant No. EP/T001062/1 via the UK Quantum Technology Hub in Computing and Simulation.

\appendix
\section{Fisher information content of measurement record}
\label{secapp:fisherinfoproof}

In this appendix, we show that that the Fisher information carried by the measurement record $\boldsymbol{h}$ regarding the amplitude $a$ is
\begin{align}
     I(a) &= \frac{N_\mathrm{shot}}{a(1-a)}\sum_{j=0}^{q-1}(2d_j + 1)^2, \nonumber 
\end{align}
as in Eq. \eqref{eq:fisher_info_predef}.

The probability of measuring a `good' state after performing the MLQAE circuit of Grover-depth $d_j$ as a function of the amplitude $a$ is
\begin{align}
    p_{d_j}(a) &= \sin^2\left[(2d_j + 1)\theta_a\right] \nonumber \\
    &= \sin^2\left[(2d_j + 1)\arcsin\sqrt{a}\right] \nonumber
\end{align}
Denoting the outcome of the single measurement as $m_{d_j}=1,0$ for a `good' or `bad' outcome respectively, the likelihood of the estimate $\tilde{a}$ being the correct amplitude $a$ can be written as 
\begin{align}
    L_{d_j}^{(\mathrm{single\ shot})}\left(\tilde{a}=a;m_{d_j}\right) &= \left[p_{d_j}(\tilde{a})\right]^{m_{d_j}}\left[1-p_{d_j}(\tilde{a})\right]^{1-m_{d_j}}. \nonumber
\end{align}
In the region around the true amplitude $a$ where the logarithm of the likelihood function is differentiable, the Fisher information gained from the measurement can be written as
\begin{align}
    I^{(\mathrm{single\ shot})}(a) &= \mathbb{E}_{m_{d_j}}\left[\left(\pdiff{}{\tilde{a}}\log L_{d_j}^{(\mathrm{single\ shot})}\right)^2 \middle| a \right] \nonumber
\end{align}
where the expectation value is being taken over the measurement outcome $m_{d_j}$ and then evaluated at the true amplitude $a$.

The derivative can be developed as
\begin{align}
    \pdiff{}{\tilde{a}}\log L_{d_j}^{(\mathrm{single\ shot})} &= \frac{\pdiff{}{\tilde{a}} L_{d_j}^{(\mathrm{single\ shot})}}{L_{d_j}^{(\mathrm{single\ shot})}} \nonumber \\
    &= \frac{\pdiff{}{\tilde{a}}\left(\left[p_{d_j}(\tilde{a})\right]^{m_{d_j}}\right)\left[1-p_{d_j}(\tilde{a})\right]^{1-m_{d_j}} + \left[p_{d_j}(\tilde{a})\right]^{m_{d_j}}\pdiff{}{\tilde{a}}\left(\left[1-p_{d_j}(\tilde{a})\right]^{1-m_{d_j}}\right)}{\left[p_{d_j}(\tilde{a})\right]^{m_{d_j}}\left[1-p_{d_j}(\tilde{a})\right]^{1-m_{d_j}}} \nonumber \\
    &= \frac{\pdiff{}{\tilde{a}}\left(\left[p_{d_j}(\tilde{a})\right]^{m_{d_j}}\right)}{\left[p_{d_j}(\tilde{a})\right]^{m_{d_j}}} + \frac{\pdiff{}{\tilde{a}}\left(\left[1-p_{d_j}(\tilde{a})\right]^{1-m_{d_j}}\right)}{\left[1-p_{d_j}(\tilde{a})\right]^{1-m_{d_j}}} \nonumber \\
    &= m_{d_j}\frac{\pdiff{}{\tilde{a}}\left[p_{d_j}(\tilde{a})\right]}{p_{d_j}(\tilde{a})} + (1-m_{d_j})\frac{\pdiff{}{\tilde{a}}\left[1-p_{d_j}(\tilde{a})\right]}{1-p_{d_j}(\tilde{a})} \nonumber
\end{align}
where the last line is valid because the $m_{d_j}$ is a binary variable.

The two remaining derivatives can be developed as
\begin{align}
    \pdiff{}{\tilde{a}}\left[p_{d_j}(\tilde{a})\right] &= \pdiff{}{\tilde{a}}\left(\sin^2\left[(2d_j + 1)\arcsin\sqrt{\tilde{a}}\right]\right) \nonumber \\
    &= 2(2d_j+1)\sin\left[(2d_j + 1)\arcsin\sqrt{\tilde{a}}\right]\cos\left[(2d_j + 1)\arcsin\sqrt{\tilde{a}}\right]\pdiff{}{\tilde{a}}\left(\arcsin\sqrt{\tilde{a}}\right) \nonumber \\
    &= \frac{2d_j+1}{\sqrt{\tilde{a}(1-\tilde{a})}}\sin\left[(2d_j + 1)\arcsin\sqrt{\tilde{a}}\right]\cos\left[(2d_j + 1)\arcsin\sqrt{\tilde{a}}\right] \nonumber \\
    &= \frac{2d_j+1}{\sqrt{\tilde{a}(1-\tilde{a})}} \sqrt{p_{d_j}(\tilde{a})\left[1-p_{d_j}(\tilde{a})\right]} \nonumber
\end{align}
and
\begin{align}
    \pdiff{}{\tilde{a}}\left[1-p_{d_j}(\tilde{a})\right] &= \pdiff{}{\tilde{a}}\left(\cos^2\left[(2d_j + 1)\arcsin\sqrt{\tilde{a}}\right]\right) \nonumber \\
    &= -2(2d_j+1)\cos\left[(2d_j + 1)\arcsin\sqrt{\tilde{a}}\right]\sin\left[(2d_j + 1)\arcsin\sqrt{\tilde{a}}\right]\pdiff{}{\tilde{a}}\left(\arcsin\sqrt{\tilde{a}}\right) \nonumber \\
    &= -\frac{2d_j+1}{\sqrt{\tilde{a}(1-\tilde{a})}}\cos\left[(2d_j + 1)\arcsin\sqrt{\tilde{a}}\right]\sin\left[(2d_j + 1)\arcsin\sqrt{\tilde{a}}\right] \nonumber \\
    &= -\frac{2d_j+1}{\sqrt{\tilde{a}(1-\tilde{a})}} \sqrt{p_{d_j}(\tilde{a})\left[1-p_{d_j}(\tilde{a})\right]}, \nonumber
\end{align}
giving 
\begin{align}
    \pdiff{}{\tilde{a}}\log L_{d_j}^{(\mathrm{single\ shot})} &= \frac{2d_j+1}{\sqrt{\tilde{a}(1-\tilde{a})}} \sqrt{p_{d_j}(\tilde{a})\left[1-p_{d_j}(\tilde{a})\right]}\left[\frac{ m_{d_j}}{p_{d_j}(\tilde{a})} - \frac{1-m_{d_j}}{1-p_{d_j}(\tilde{a})}\right] \nonumber\\
    & = \frac{\left(2d_j+1\right)\left(m_{d_j}\left[1-p_{d_j}(\tilde{a})\right] - \left[1-m_{d_j}\right]\left[p_{d_j}(\tilde{a})\right] \right)}{\sqrt{\tilde{a}(1-\tilde{a})\left[p_{d_j}(\tilde{a})\right]\left[1-p_{d_j}(\tilde{a})\right]}}. \nonumber
\end{align}
As $m_{d_j}$ is a binary value, squaring this gives
\begin{align}
    \left(\pdiff{}{\tilde{a}}\log L_{d_j}^{(\mathrm{single\ shot})}\right)^2 &= \frac{\left(2d_j+1\right)^2\left(m_{d_j}\left[1-p_{d_j}(\tilde{a})\right]^2 + \left[1-m_{d_j}\right]\left[p_{d_j}(\tilde{a})\right]^2 \right)}{\tilde{a}(1-\tilde{a})\left[p_{d_j}(\tilde{a})\right]\left[1-p_{d_j}(\tilde{a})\right]}, \nonumber
\end{align}
and, finally, the expectation value over $m_{d_j}$ can be taken and evaluated at the true amplitude $a$ to find the Fisher information as
\begin{align}
    I^{(\mathrm{single\ shot})}(a) &= \frac{\left(2d_j+1\right)^2\left(\left[p_{d_j}(a)\right]\left[1-p_{d_j}(a)\right]^2 + \left[1-p_{d_j}(a)\right]\left[p_{d_j}(a)\right]^2 \right)}{a(1-a)\left[p_{d_j}(a)\right]\left[1-p_{d_j}(a)\right]} \nonumber \\
    &= \frac{(2d_j+1)^2}{a(1-a)}\left(\left[1-p_{d_j}(a)\right] + \left[p_{d_j}(a)\right]\right) \nonumber \\
    &= \frac{(2d_j+1)^2}{a(1-a)}. \nonumber
\end{align}

As Fisher information is additive, the single shot Fisher information $I^{\mathrm{single\ shot}}(a)$ can be summed over the $N_\mathrm{shot}$ shots at each Grover-depth $d_j$ to give the total fisher information 
\begin{align}
     I(a) &= \frac{N_\mathrm{shot}}{a(1-a)}\sum_{j=0}^{q-1}(2d_j + 1)^2, \nonumber 
\end{align}
as required.

\section{Analysis of the base for the exponential schedule}
\label{secapp:nubase}

In subsection \ref{ssec:depthlimited}, we defined a slight modification to the exponential Grover-depth schedule $D_\mathrm{EXP}$ (from Eq. \eqref{eq:expsched}) to accommodate a maximum Grover-depth $d$ that is not a power of $2$.
The modification, as stated in Eq. \eqref{eq:expschednu}, is given by 
\begin{align}
    D_{\mathrm{EXP},\nu} &= \left\{d_0 = 0\right\}\cup \left\{d_j = \mathrm{Round}\left(\nu^{j-1}\right)\right\}_{j=1}^{q-1}, \nonumber 
\end{align}
where the base $\nu$ and number of depths $q$ are chosen together such that $\nu$ is the closest value to $2$ such that $\nu^{q-2}=d$.
However, our scaling analyses consider only the case where the maximum Grover-depth $d$ is a power of 2; that is, for $\nu=2$.
We assume that the scaling results still apply for other values of the base $\nu$ and justify this with the claim  that ``the difference $|\nu-2|$ becomes smaller over larger scales of maximum Grover-depth $d$".
In this appendix, we briefly show that this is true.

By definition,
\begin{align}
    \nu_{\mathrm{lower}} < \nu < \nu_{\mathrm{upper}},\nonumber
\end{align}
where
\begin{align}
    \nu_\mathrm{lower}^{q-2} &= 2^{q-3} \nonumber\\
    \nu_\mathrm{upper}^{q-2} &= 2^{q-1}. \nonumber
\end{align}
These bounds can be evaluated as
\begin{align}
    \nu_\mathrm{lower} &= 2^{\frac{q-3}{q-2}}\nonumber\\
    \nu_\mathrm{upper} &= 2^{\frac{q-1}{q-2}},\nonumber
\end{align}
both of which tend to $2$ with increasing $q$, and thus with increasing Grover-depth $d$.

\bibliography{main}

\begin{thebibliography}{50}
\providecommand{\natexlab}[1]{#1}
\providecommand{\url}[1]{\texttt{#1}}
\expandafter\ifx\csname urlstyle\endcsname\relax
  \providecommand{\doi}[1]{doi: #1}\else
  \providecommand{\doi}{doi: \begingroup \urlstyle{rm}\Url}\fi

\bibitem[Montanaro(2016)]{montanaro2016quantum}
Ashley Montanaro.
\newblock Quantum algorithms: an overview.
\newblock \emph{npj Quantum Information}, 2\penalty0 (1):\penalty0 15023, 2016.
\newblock \doi{10.1038/npjqi.2015.23}.
\newblock URL \url{https://doi.org/10.1038/npjqi.2015.23}.

\bibitem[Cho et~al.(2021)Cho, Chen, Chen, Huang, Hsu, Cao, Zeng, Tan, and
  Chang]{cho2021quantum}
Chien-Hung Cho, Chih-Yu Chen, Kuo-Chin Chen, Tsung-Wei Huang, Ming-Chien Hsu,
  Ning-Ping Cao, Bei Zeng, Seng-Ghee Tan, and Ching-Ray Chang.
\newblock {Quantum computation: Algorithms and Applications}.
\newblock \emph{Chinese Journal of Physics}, 72:\penalty0 248--269, 2021.
\newblock \doi{https://doi.org/10.1016/j.cjph.2021.05.001}.
\newblock URL
  \url{https://www.sciencedirect.com/science/article/pii/S0577907321001039}.

\bibitem[Feynman(1982)]{feynman1982simulating}
Richard~P. Feynman.
\newblock Simulating physics with computers.
\newblock \emph{International Journal of Theoretical Physics}, 21\penalty0
  (6):\penalty0 467--488, 1982.
\newblock \doi{10.1007/BF02650179}.
\newblock URL \url{https://doi.org/10.1007/BF02650179}.

\bibitem[Benioff(1980)]{benioff1980computer}
Paul Benioff.
\newblock {The computer as a physical system: A microscopic quantum mechanical
  Hamiltonian model of computers as represented by Turing machines}.
\newblock \emph{Journal of statistical physics}, 22\penalty0 (5):\penalty0
  563--591, 1980.
\newblock URL \url{https://doi.org/10.1007/BF01011339}.

\bibitem[Deutsch and Penrose(1985)]{deutsch1985quantum}
David Deutsch and Roger Penrose.
\newblock {Quantum theory, the Church-Turing principle and the universal
  quantum computer}.
\newblock \emph{Proceedings of the Royal Society of London. A. Mathematical and
  Physical Sciences}, 400\penalty0 (1818):\penalty0 97--117, 1985.
\newblock \doi{10.1098/rspa.1985.0070}.
\newblock URL
  \url{https://royalsocietypublishing.org/doi/abs/10.1098/rspa.1985.0070}.

\bibitem[Montanaro(2015)]{montanaro2015quantum}
Ashley Montanaro.
\newblock {Quantum speedup of Monte Carlo methods}.
\newblock \emph{Proceedings of the Royal Society A: Mathematical, Physical and
  Engineering Sciences}, 471\penalty0 (2181):\penalty0 20150301, 2015.
\newblock \doi{10.1098/rspa.2015.0301}.
\newblock URL
  \url{https://royalsocietypublishing.org/doi/abs/10.1098/rspa.2015.0301}.

\bibitem[Rebentrost et~al.(2018)Rebentrost, Gupt, and
  Bromley]{rebenstrost2018quantum}
Patrick Rebentrost, Brajesh Gupt, and Thomas~R. Bromley.
\newblock {Quantum computational finance: Monte Carlo pricing of financial
  derivatives}.
\newblock \emph{Phys. Rev. A}, 98:\penalty0 022321, 2018.
\newblock \doi{10.1103/PhysRevA.98.022321}.
\newblock URL \url{https://link.aps.org/doi/10.1103/PhysRevA.98.022321}.

\bibitem[Miyamoto and Shiohara(2020)]{miyamoto2020reduction}
Koichi Miyamoto and Kenji Shiohara.
\newblock {Reduction of qubits in a quantum algorithm for Monte Carlo
  simulation by a pseudo-random-number generator}.
\newblock \emph{Phys. Rev. A}, 102:\penalty0 022424, 2020.
\newblock \doi{10.1103/PhysRevA.102.022424}.
\newblock URL \url{https://link.aps.org/doi/10.1103/PhysRevA.102.022424}.

\bibitem[Wiebe et~al.(2015)Wiebe, Kapoor, and M.~Svore]{wiebe2015quantum}
Nathan Wiebe, Ashish Kapoor, and Krysta M.~Svore.
\newblock Quantum algorithms for nearest-neighbor methods for supervised and
  unsupervised learning.
\newblock \emph{Quantum Info. Comput.}, 15\penalty0 (3–4):\penalty0 316--356,
  2015.
\newblock URL \url{https://doi.org/10.26421/QIC15.3-4-7}.

\bibitem[Wiebe et~al.(2016{\natexlab{a}})Wiebe, Kapoor, and
  M.~Svore]{wiebe2016quantum}
Nathan Wiebe, Ashish Kapoor, and Krysta M.~Svore.
\newblock Quantum deep learning.
\newblock \emph{Quantum Info. Comput.}, 16\penalty0 (7–8):\penalty0 541--587,
  2016{\natexlab{a}}.
\newblock URL \url{https://doi.org/10.26421/QIC16.7-8-1}.

\bibitem[Wiebe et~al.(2016{\natexlab{b}})Wiebe, Kapoor, and
  Svore]{wiebe2016quantum2}
Nathan Wiebe, Ashish Kapoor, and Krysta~M Svore.
\newblock {Quantum Perceptron Models}.
\newblock In \emph{Proceedings of the 30th International Conference on Neural
  Information Processing Systems}, NIPS'16, page 4006–4014, Red Hook, NY,
  USA, 2016{\natexlab{b}}. Curran Associates Inc.
\newblock URL \url{https://dl.acm.org/doi/10.5555/3157382.3157545}.

\bibitem[Li et~al.(2019)Li, Chakrabarti, and Wu]{li2019sublinear}
Tongyang Li, Shouvanik Chakrabarti, and Xiaodi Wu.
\newblock Sublinear quantum algorithms for training linear and kernel-based
  classifiers.
\newblock In Kamalika Chaudhuri and Ruslan Salakhutdinov, editors,
  \emph{Proceedings of the 36th International Conference on Machine Learning},
  volume~97 of \emph{Proceedings of Machine Learning Research}, pages
  3815--3824. PMLR, 2019.
\newblock URL \url{https://proceedings.mlr.press/v97/li19b.html}.

\bibitem[Kerenidis et~al.(2019)Kerenidis, Landman, Luongo, and
  Prakash]{kerenidis2019qmeans}
Iordanis Kerenidis, Jonas Landman, Alessandro Luongo, and Anupam Prakash.
\newblock q-means: A quantum algorithm for unsupervised machine learning.
\newblock In H.~Wallach, H.~Larochelle, A.~Beygelzimer, F.~d\textquotesingle
  Alch\'{e}-Buc, E.~Fox, and R.~Garnett, editors, \emph{{Advances in Neural
  Information Processing Systems}}, volume~32. Curran Associates, Inc., 2019.
\newblock URL
  \url{https://proceedings.neurips.cc/paper/2019/hash/16026d60ff9b54410b3435b403afd226-Abstract.html}.

\bibitem[Miyahara et~al.(2020)Miyahara, Aihara, and
  Lechner]{miyahara2020quantum}
Hideyuki Miyahara, Kazuyuki Aihara, and Wolfgang Lechner.
\newblock Quantum expectation-maximization algorithm.
\newblock \emph{Phys. Rev. A}, 101:\penalty0 012326, 2020.
\newblock \doi{10.1103/PhysRevA.101.012326}.
\newblock URL \url{https://link.aps.org/doi/10.1103/PhysRevA.101.012326}.

\bibitem[Brassard et~al.(2002)Brassard, Høyer, Mosca, and
  Tapp]{brassard2002quantum}
Gilles Brassard, Peter Høyer, Michele Mosca, and Alain Tapp.
\newblock {Quantum Amplitude Amplification and Estimation}.
\newblock \emph{Contemp. Math. Ser. Millenn.}, 305:\penalty0 53--74, 2002.
\newblock \doi{10.1090/conm/305}.
\newblock URL \url{http://dx.doi.org/10.1090/conm/305}.

\bibitem[Grover(1996)]{grover1996fast}
Lov~K Grover.
\newblock A fast quantum mechanical algorithm for database search.
\newblock In \emph{{Proceedings of the twenty-eighth annual ACM symposium on
  Theory of computing}}, pages 212--219, 1996.
\newblock URL \url{https://dl.acm.org/doi/pdf/10.1145/237814.237866}.

\bibitem[Kitaev(1995)]{kitaev1995quantum}
A~Yu Kitaev.
\newblock {Quantum measurements and the Abelian stabilizer problem}, 1995.
\newblock URL \url{https://arxiv.org/abs/quant-ph/9511026}.
\newblock arXiv preprint quant-ph/9511026.

\bibitem[Aaronson and Rall(2020)]{aaronson2020quantum}
Scott Aaronson and Patrick Rall.
\newblock \emph{{Quantum Approximate Counting, Simplified}}, pages 24--32.
\newblock SIAM, 2020.
\newblock \doi{10.1137/1.9781611976014.5}.
\newblock URL \url{https://doi.org/10.1137/1.9781611976014.5}.

\bibitem[Venkateswaran and O'Donnell(2021)]{venkateswaran2020quantum}
Ramgopal Venkateswaran and Ryan O'Donnell.
\newblock {Quantum approximate counting with nonadaptive Grover iterations}.
\newblock In \emph{{38th International Symposium on Theoretical Aspects of
  Computer Science}}, pages 59:1--59:12, 2021.
\newblock URL \url{https://arxiv.org/pdf/2010.04370.pdf}.

\bibitem[Nakaji(2020)]{nakaji2020faster}
Kouhei Nakaji.
\newblock Faster amplitude estimation.
\newblock \emph{Quantum Info. Comput.}, 20\penalty0 (13–14):\penalty0
  1109--1123, 2020.
\newblock URL \url{https://doi.org/10.26421/QIC20.13-14-2}.

\bibitem[Grinko et~al.(2021)Grinko, Gacon, Zoufal, and
  Woerner]{grinko2021iterative}
Dmitry Grinko, Julien Gacon, Christa Zoufal, and Stefan Woerner.
\newblock Iterative quantum amplitude estimation.
\newblock \emph{npj Quantum Information}, 7\penalty0 (1):\penalty0 52, 2021.
\newblock \doi{10.1038/s41534-021-00379-1}.
\newblock URL \url{https://doi.org/10.1038/s41534-021-00379-1}.

\bibitem[Rall(2021)]{rall2021faster}
Patrick Rall.
\newblock Faster {C}oherent {Q}uantum {A}lgorithms for {P}hase, {E}nergy, and
  {A}mplitude {E}stimation.
\newblock \emph{{Quantum}}, 5:\penalty0 566, 2021.
\newblock \doi{10.22331/q-2021-10-19-566}.
\newblock URL \url{https://doi.org/10.22331/q-2021-10-19-566}.

\bibitem[Rall and Fuller(2022)]{rall2022amplitude}
Patrick Rall and Bryce Fuller.
\newblock {Amplitude Estimation from Quantum Signal Processing}, 2022.
\newblock URL \url{https://arxiv.org/abs/2207.08628}.
\newblock arXiv preprint arXiv:2207.08628.

\bibitem[Manzano et~al.(2022)Manzano, Musso, and Leitao]{manzano2022real}
Alberto Manzano, Daniele Musso, and {\'A}lvaro Leitao.
\newblock Real quantum amplitude estimation, 2022.
\newblock URL \url{https://arxiv.org/abs/2204.13641}.
\newblock arXiv preprint arXiv:2204.13641.

\bibitem[Giurgica-Tiron et~al.(2022{\natexlab{a}})Giurgica-Tiron, Kerenidis,
  Labib, Prakash, and Zeng]{giurgicatiron2022low}
Tudor Giurgica-Tiron, Iordanis Kerenidis, Farrokh Labib, Anupam Prakash, and
  William Zeng.
\newblock Low depth algorithms for quantum amplitude estimation.
\newblock \emph{{Quantum}}, 6:\penalty0 745, 2022{\natexlab{a}}.
\newblock \doi{10.22331/q-2022-06-27-745}.
\newblock URL \url{https://doi.org/10.22331/q-2022-06-27-745}.

\bibitem[Fukuzawa et~al.(2022)Fukuzawa, Ho, Irani, and
  Zion]{fukuzawa2022modified}
Shion Fukuzawa, Christopher Ho, Sandy Irani, and Jasen Zion.
\newblock {Modified Iterative Quantum Amplitude Estimation is Asymptotically
  Optimal}, 2022.
\newblock URL \url{https://arxiv.org/abs/2208.14612}.
\newblock arXiv preprint arXiv:2208.14612.

\bibitem[Suzuki et~al.(2020)Suzuki, Uno, Raymond, Tanaka, Onodera, and
  Yamamoto]{suzuki2020amplitude}
Yohichi Suzuki, Shumpei Uno, Rudy Raymond, Tomoki Tanaka, Tamiya Onodera, and
  Naoki Yamamoto.
\newblock Amplitude estimation without phase estimation.
\newblock \emph{Quantum Information Processing}, 19\penalty0 (2):\penalty0 75,
  2020.
\newblock \doi{10.1007/s11128-019-2565-2}.
\newblock URL \url{https://doi.org/10.1007/s11128-019-2565-2}.

\bibitem[Brown et~al.(2020)Brown, Goktas, and Tham]{brown2020quantum}
Eric~G Brown, Oktay Goktas, and WK~Tham.
\newblock Quantum amplitude estimation in the presence of noise, 2020.
\newblock URL \url{https://arxiv.org/abs/2006.14145}.
\newblock arXiv preprint arXiv:2006.14145.

\bibitem[Tanaka et~al.(2021)Tanaka, Suzuki, Uno, Raymond, Onodera, and
  Yamamoto]{tanaka2021amplitude}
Tomoki Tanaka, Yohichi Suzuki, Shumpei Uno, Rudy Raymond, Tamiya Onodera, and
  Naoki Yamamoto.
\newblock Amplitude estimation via maximum likelihood on noisy quantum
  computer.
\newblock \emph{Quantum Information Processing}, 20\penalty0 (9):\penalty0 293,
  2021.
\newblock \doi{10.1007/s11128-021-03215-9}.
\newblock URL \url{https://doi.org/10.1007/s11128-021-03215-9}.

\bibitem[Herbert et~al.(2021)Herbert, Guichard, and Ng]{herbert2021noise}
Steven Herbert, Roland Guichard, and Darren Ng.
\newblock {Noise-Aware Quantum Amplitude Estimation}, 2021.
\newblock URL \url{https://arxiv.org/abs/2109.04840}.
\newblock arXiv preprint arXiv:2109.04840.

\bibitem[Uno et~al.(2021)Uno, Suzuki, Hisanaga, Raymond, Tanaka, Onodera, and
  Yamamoto]{uno2021modified}
Shumpei Uno, Yohichi Suzuki, Keigo Hisanaga, Rudy Raymond, Tomoki Tanaka,
  Tamiya Onodera, and Naoki Yamamoto.
\newblock {Modified Grover operator for quantum amplitude estimation}.
\newblock \emph{New Journal of Physics}, 23\penalty0 (8):\penalty0 083031,
  2021.
\newblock \doi{10.1088/1367-2630/ac19da}.
\newblock URL \url{https://doi.org/10.1088/1367-2630/ac19da}.

\bibitem[Tanaka et~al.(2022)Tanaka, Uno, Onodera, Yamamoto, and
  Suzuki]{tanaka2022noisy}
Tomoki Tanaka, Shumpei Uno, Tamiya Onodera, Naoki Yamamoto, and Yohichi Suzuki.
\newblock Noisy quantum amplitude estimation without noise estimation.
\newblock \emph{Phys. Rev. A}, 105:\penalty0 012411, 2022.
\newblock \doi{10.1103/PhysRevA.105.012411}.
\newblock URL \url{https://link.aps.org/doi/10.1103/PhysRevA.105.012411}.

\bibitem[Plekhanov et~al.(2022)Plekhanov, Rosenkranz, Fiorentini, and
  Lubasch]{plekhanov2022variational}
Kirill Plekhanov, Matthias Rosenkranz, Mattia Fiorentini, and Michael Lubasch.
\newblock Variational quantum amplitude estimation.
\newblock \emph{{Quantum}}, 6:\penalty0 670, 2022.
\newblock \doi{10.22331/q-2022-03-17-670}.
\newblock URL \url{https://doi.org/10.22331/q-2022-03-17-670}.

\bibitem[Callison and Chancellor(2022)]{callison2022hybrid}
Adam Callison and Nicholas Chancellor.
\newblock Hybrid quantum-classical algorithms in the noisy intermediate-scale
  quantum era and beyond.
\newblock \emph{Phys. Rev. A}, 106:\penalty0 010101, 2022.
\newblock \doi{10.1103/PhysRevA.106.010101}.
\newblock URL \url{https://link.aps.org/doi/10.1103/PhysRevA.106.010101}.

\bibitem[van~der Vaart(1998)]{vandervaart1998bernstein}
A.~W. van~der Vaart.
\newblock {Bernstein-von Mises Theorem}.
\newblock In \emph{{Asymptotic Statistics}}, chapter 10.2, pages 140--146.
  Cambride University Press, Cambridge, UK, 1998.

\bibitem[Higgins et~al.(2009)Higgins, Berry, Bartlett, Mitchell, Wiseman, and
  Pryde]{higgins2009demonstrating}
B~L Higgins, D~W Berry, S~D Bartlett, M~W Mitchell, H~M Wiseman, and G~J Pryde.
\newblock Demonstrating heisenberg-limited unambiguous phase estimation without
  adaptive measurements.
\newblock \emph{New Journal of Physics}, 11\penalty0 (7):\penalty0 073023,
  2009.
\newblock \doi{10.1088/1367-2630/11/7/073023}.
\newblock URL \url{https://dx.doi.org/10.1088/1367-2630/11/7/073023}.

\bibitem[Kimmel et~al.(2015)Kimmel, Low, and Yoder]{kimmel2015robust}
Shelby Kimmel, Guang~Hao Low, and Theodore~J. Yoder.
\newblock Robust calibration of a universal single-qubit gate set via robust
  phase estimation.
\newblock \emph{Phys. Rev. A}, 92:\penalty0 062315, Dec 2015.
\newblock \doi{10.1103/PhysRevA.92.062315}.
\newblock URL \url{https://link.aps.org/doi/10.1103/PhysRevA.92.062315}.

\bibitem[Wiebe and Granade(2016)]{wiebe2016efficient}
Nathan Wiebe and Chris Granade.
\newblock {Efficient Bayesian Phase Estimation}.
\newblock \emph{Phys. Rev. Lett.}, 117:\penalty0 010503, 2016.
\newblock \doi{10.1103/PhysRevLett.117.010503}.
\newblock URL \url{https://link.aps.org/doi/10.1103/PhysRevLett.117.010503}.

\bibitem[Knill et~al.(2007)Knill, Ortiz, and Somma]{knill2007optimal}
Emanuel Knill, Gerardo Ortiz, and Rolando~D. Somma.
\newblock Optimal quantum measurements of expectation values of observables.
\newblock \emph{Phys. Rev. A}, 75:\penalty0 012328, 2007.
\newblock \doi{10.1103/PhysRevA.75.012328}.
\newblock URL \url{https://link.aps.org/doi/10.1103/PhysRevA.75.012328}.

\bibitem[Dutkiewicz et~al.(2021)Dutkiewicz, Terhal, and
  O'Brien]{dutkiewicz2021heisenberg}
Alicja Dutkiewicz, Barbara~M Terhal, and Thomas~E O'Brien.
\newblock {Heisenberg-limited quantum phase estimation of multiple eigenvalues
  with few control qubits}, 2021.
\newblock URL \url{https://arxiv.org/abs/2107.04605}.
\newblock arXiv preprint arXiv:2107.04605.

\bibitem[Zintchenko and Wiebe(2016)]{zintchenko2016randomized}
Ilia Zintchenko and Nathan Wiebe.
\newblock Randomized gap and amplitude estimation.
\newblock \emph{Phys. Rev. A}, 93:\penalty0 062306, 2016.
\newblock \doi{10.1103/PhysRevA.93.062306}.
\newblock URL \url{https://link.aps.org/doi/10.1103/PhysRevA.93.062306}.

\bibitem[Venegas-Andraca(2012)]{venegasandraca2012quantum}
Salvador~El{\'i}as Venegas-Andraca.
\newblock Quantum walks: a comprehensive review.
\newblock \emph{Quantum Information Processing}, 11\penalty0 (5):\penalty0
  1015--1106, 2012.
\newblock \doi{10.1007/s11128-012-0432-5}.
\newblock URL \url{https://doi.org/10.1007/s11128-012-0432-5}.

\bibitem[Egger et~al.(2020)Egger, Gambella, Marecek, McFaddin, Mevissen,
  Raymond, Simonetto, Woerner, and Yndurain]{egger2020quantum}
Daniel~J. Egger, Claudio Gambella, Jakub Marecek, Scott McFaddin, Martin
  Mevissen, Rudy Raymond, Andrea Simonetto, Stefan Woerner, and Elena Yndurain.
\newblock {Quantum Computing for Finance: State-of-the-Art and Future
  Prospects}.
\newblock \emph{IEEE Transactions on Quantum Engineering}, 1:\penalty0 1--24,
  2020.
\newblock \doi{10.1109/TQE.2020.3030314}.
\newblock URL \url{https://doi.org/10.1109/TQE.2020.3030314}.

\bibitem[Egger et~al.(2021)Egger, García~Gutiérrez, Mestre, and
  Woerner]{egger2021credit}
Daniel~J. Egger, Ricardo García~Gutiérrez, Jordi~Cahué Mestre, and Stefan
  Woerner.
\newblock Credit risk analysis using quantum computers.
\newblock \emph{IEEE Transactions on Computers}, 70\penalty0 (12):\penalty0
  2136--2145, 2021.
\newblock \doi{10.1109/TC.2020.3038063}.
\newblock URL \url{https://doi.org/10.1109/TQE.2020.3030314}.

\bibitem[Giurgica-Tiron et~al.(2022{\natexlab{b}})Giurgica-Tiron, Johri,
  Kerenidis, Nguyen, Pisenti, Prakash, Sosnova, Wright, and
  Zeng]{giurgicatiron2022low2}
Tudor Giurgica-Tiron, Sonika Johri, Iordanis Kerenidis, Jason Nguyen, Neal
  Pisenti, Anupam Prakash, Ksenia Sosnova, Ken Wright, and William Zeng.
\newblock Low-depth amplitude estimation on a trapped-ion quantum computer.
\newblock \emph{Phys. Rev. Research}, 4:\penalty0 033034, 2022{\natexlab{b}}.
\newblock \doi{10.1103/PhysRevResearch.4.033034}.
\newblock URL \url{https://link.aps.org/doi/10.1103/PhysRevResearch.4.033034}.

\bibitem[Wie(2019)]{wie2019simpler}
Chu-Ryang Wie.
\newblock Simpler quantum counting.
\newblock \emph{Quantum Info. Comput.}, 19\penalty0 (11–12):\penalty0
  0967--0983, 2019.
\newblock URL \url{https://doi.org/10.26421/QIC19.11-12-5}.

\bibitem[Wang et~al.(2021)Wang, Koh, Johnson, and Cao]{wang2021minimizing}
Guoming Wang, Dax~Enshan Koh, Peter~D. Johnson, and Yudong Cao.
\newblock Minimizing estimation runtime on noisy quantum computers.
\newblock \emph{PRX Quantum}, 2:\penalty0 010346, 2021.
\newblock \doi{10.1103/PRXQuantum.2.010346}.
\newblock URL \url{https://link.aps.org/doi/10.1103/PRXQuantum.2.010346}.

\bibitem[Rao et~al.(2020)Rao, Yu, Lim, Jin, and Choi]{rao2020quantum}
Pooja Rao, Kwangmin Yu, Hyunkyung Lim, Dasol Jin, and Deokkyu Choi.
\newblock {Quantum amplitude estimation algorithms on IBM quantum devices}.
\newblock In Keith~S. Deacon, editor, \emph{Quantum Communications and Quantum
  Imaging XVIII}, volume 11507, pages 49 -- 60. International Society for
  Optics and Photonics, SPIE, 2020.
\newblock \doi{10.1117/12.2568748}.
\newblock URL \url{https://doi.org/10.1117/12.2568748}.

\bibitem[Blair et~al.(1976)Blair, Edwards, and Johnson]{blair1976rational}
JM~Blair, CA~Edwards, and J~Howard Johnson.
\newblock {Rational Chebyshev approximations for the inverse of the error
  function}.
\newblock \emph{Mathematics of Computation}, 30\penalty0 (136):\penalty0
  827--830, 1976.
\newblock \doi{10.1090/S0025-5718-1976-0421040-7}.
\newblock URL \url{http://dx.doi.org/10.1090/S0025-5718-1976-0421040-7}.

\bibitem[Babbush et~al.(2021)Babbush, McClean, Newman, Gidney, Boixo, and
  Neven]{babbush2021focus}
Ryan Babbush, Jarrod~R. McClean, Michael Newman, Craig Gidney, Sergio Boixo,
  and Hartmut Neven.
\newblock {Focus beyond Quadratic Speedups for Error-Corrected Quantum
  Advantage}.
\newblock \emph{PRX Quantum}, 2:\penalty0 010103, 2021.
\newblock \doi{10.1103/PRXQuantum.2.010103}.
\newblock URL \url{https://link.aps.org/doi/10.1103/PRXQuantum.2.010103}.

\end{thebibliography}

\end{document}